\documentclass[a4paper,fleqn]{mnras}

\usepackage{amsmath,amssymb}
\usepackage{txfonts}

\usepackage[T1]{fontenc}
\usepackage{ae,aecompl}

\usepackage{natbib}
\usepackage{url}
\usepackage{graphicx}
\bibpunct[; ]{(}{)}{;}{a}{}{,}

\newcommand{\hatlas}{\mbox{H-ATLAS}}
\newcommand{\sigmapos}{\ensuremath{\sigma_\text{pos}}}
\newcommand{\mum}{\ensuremath{\mu}\text{m}}
\newcommand{\degree}{\ensuremath{^\circ}}
\newcommand{\Herschel}{\emph{Herschel}}
\newcommand{\hour}{\ensuremath{^\text{h}}}
\newcommand{\eqnref}[1]{equation~(\ref{#1})}

\defcitealias{Smith2011a}{S11}
\defcitealias{Valiante2014}{Paper I}

\title[Multi-wavelength IDs for H-ATLAS]{The \Herschel\footnotemark-ATLAS Data Release 1 Paper II: Multi-wavelength counterparts to submillimetre sources}
\author[N. Bourne et al.]%
  {N.~Bourne,$^{1}$\footnotemark
   L.~Dunne,$^{1,2}$
   S.\,J.~Maddox,$^{1,2}$
   S.~Dye,$^3$
   C.~Furlanetto,$^{3,4}$\newauthor
   C.~Hoyos,$^{3,5,6}$
   D.\,J.\,B.~Smith,$^7$
   S.~Eales,$^2$
   M.\,W.\,L.~Smith,$^2$  
   E.~Valiante,$^2$\newauthor
   M. Alpaslan,$^{8,9}$
   E. Andrae,$^{10}$
   I.\,K.~Baldry,$^{11}$
   M.\,E. Cluver,$^{12}$
   A. Cooray,$^{13}$ \newauthor
   S.\,P.~Driver,$^{8,9}$ 
   J.\,S.~Dunlop,$^1$ 
   M.\,W.~Grootes,$^{10}$
   R.\,J.~Ivison,$^{1,14}$
   T.\,H. Jarrett,$^{15}$  \newauthor
   J.~Liske,$^{16}$ 
   B.\,F. Madore,$^{17}$ 
   C.\,C. Popescu,$^{18}$ 
   A.\,G.~Robotham,$^{8,9}$ 
   K.~Rowlands,$^{8}$ \newauthor
   M. Seibert,$^{17}$
   M.\,A. Thompson,$^{7}$
   R.\,J. Tuffs,$^{10}$
   S. Viaene,$^{19}$
   A.\,H. Wright$^{9}$
   \\
   $^1$ SUPA\footnotemark, Institute for Astronomy, University of Edinburgh, Royal Observatory, Edinburgh, EH9 3HJ, UK\\
   $^2$ School of Physics and Astronomy, Cardiff University, The Parade, Cardiff, CF24 3AA, UK\\
   $^3$ School of Physics and Astronomy, University of Nottingham, University Park, Nottingham, NG7 2RD, UK\\
   $^4$ CAPES Foundation, Ministry of Education of Brazil, Bras\'ilia/DF, 70040-020, Brazil\\
   $^5$ Departamento de Física Teórica, Universidad Autónoma de Madrid, Crta de Colmenar Viejo. km 15.600. 28049 Madrid, Spain\\
   $^6$ Departamento de Física, Universidad Carlos III de Madrid, Campus de Leganés, Avda. Universidad 30. 28911 Madrid, Spain\\   
   $^7$ Centre for Astrophysics Research, School of Physics, Astronomy and Mathematics, University of Hertfordshire, College Lane, \\ Hatfield, AL10 9AB, UK\\
   $^{8}$ SUPA, School of Physics and Astronomy, University of St Andrews, North Haugh, St Andrews, Fife, KY16 9SS, UK\\
   $^{9}$ International Centre for Radio Astronomy (ICRAR), University of Western Australia, Crawley, WA6009, Australia\\
   $^{10}$ Max-Planck-Institut fuer Kernphysik (MPIK), Saupfercheckweg 1, 69117 Heidelberg, Germany\\
   $^{11}$ Astrophysics Research Institute, Liverpool John Moores University, IC2, Liverpool Science Park, 146 Brownlow Hill, Liverpool, \\ L3 5RF, UK\\
   $^{12}$ University of the Western Cape, Robert Sobukwe Road, Bellville, 7535, South Africa\\
   $^{13}$ Department of Physics and Astronomy, University of California, Irvine, CA 92697\\
   $^{14}$ ESO, Karl Schwarzschild Strasse 2, D-85748 Garching, Germany\\
   $^{15}$ Department of Astronomy, University of Cape Town, Private Bag X3, Rondebosch, 7701, South Africa\\
   $^{16}$ Hamburger Sternwarte, Universit{\"a}t Hamburg, Gojenbergsweg 112, 21029 Hamburg, Germany\\
   $^{17}$ The Observatories, Carnegie Institute of Washington, 813 Santa Barbara Street, Pasadena, CA 91101\\
   $^{18}$ Jeremiah Horrocks Institute, University of Central Lancashire, Preston, PR1 2HE, UK\\
   $^{19}$ Sterrenkundig Observatorium, Universiteit Gent, Krijgslaan 281, B-9000 Gent, Belgium
}
\date{Accepted XXX. Received YYY; in original form ZZZ}

\pubyear{2016}

\begin{document}
\label{firstpage}
\pagerange{\pageref{firstpage}--\pageref{lastpage}}
\maketitle

\label{firstpage}

\maketitle

\begin{abstract}
This paper is the second in a pair of articles presenting data release 1 (DR1) of the \Herschel\ Astrophysical Terahertz Large Area Survey (\hatlas), the largest single open-time key project carried out with the \Herschel\ Space Observatory.
The \hatlas\ is a wide-area imaging survey carried out in five photometric bands at 100, 160, 250, 350 and 500\mum\ covering a total area of 600\,deg$^2$.
In this paper we describe the identification of optical counterparts to submillimetre sources in DR1, comprising an area of 161~deg$^2$ over three equatorial fields of roughly $12\degree\times4.5\degree$ centred at 9\hour, 12\hour\ and 14.5\hour\ respectively.
Of all the \hatlas\ fields, the equatorial regions benefit from the greatest overlap with current multi-wavelength surveys 
spanning ultraviolet (UV) to mid-infrared regimes, as well as extensive spectroscopic coverage.
We use a likelihood-ratio technique to identify SDSS counterparts at $r<22.4$ for 250-\mum-selected sources detected at $\geq 4\sigma$ ($\approx$28\,mJy).
We find `reliable' counterparts (reliability $R\geq0.8$) for 44,835 sources (39 per cent), with an estimated completeness of 73.0 per cent and contamination rate of 4.7 per cent.
Using redshifts and multi-wavelength photometry from GAMA and other public catalogues, we show that \hatlas-selected galaxies at $z<0.5$ span a wide range of optical colours, total infrared (IR) luminosities, and IR/UV ratios, with no strong disposition towards mid-IR-classified AGN in comparison with optical selection.
The data described herein, together with all maps and catalogues described in the companion paper \citep{Valiante2014}, are available from the \hatlas\ website at \url{www.h-atlas.org}.
\end{abstract}


\section{Introduction}
\label{sec:intro}

\renewcommand{\thefootnote}{\fnsymbol{footnote}}
\footnotetext[1]{\Herschel\ is an ESA space observatory with science instruments provided by European-led Principal Investigator consortia and with important participation from NASA.}
\footnotetext[2]{nbourne22@gmail.com}
\footnotetext[3]{Scottish Universities Physics Alliance}
\renewcommand{\thefootnote}{\arabic{footnote}}

The \Herschel\ Astrophysical Large Area Survey (\hatlas; \citealp{Eales2010a}) is the largest-area submillimetre (submm) survey conducted with the \Herschel\ Space Observatory \citep{Pilbratt2010}, imaging around 600~deg$^2$ at 100 and 160\mum\ with PACS \citep{Poglitsch2010} and 250, 350 and 500\mum\ with SPIRE \citep{Griffin2010}, as described by \citet{Ibar2010}, \citet{Pascale2010} and \citet[hereafter \citetalias{Valiante2014}]{Valiante2014}.
Data release 1(DR1) covers 161~deg$^2$ in three equatorial fields at RA of approximately 9\hour, 12\hour\ and 14.5\hour\ (GAMA9, GAMA12, GAMA15), and benefits from extensive multi-wavelength coverage in the Sloan Digital Sky Survey (SDSS; \citealp{York2000}), the \emph{Galaxy Evolution Explorer} (\emph{GALEX}; \citealp{Martin2005}), the UK Infrared Deep Sky Survey Large Area Survey (UKIDSS-LAS; \citealp{Lawrence2007}), the \emph{Wide-field Infrared Survey Explorer} (\emph{WISE}; \citealp{Wright2010}), the VISTA Kilo-degree Infrared Galaxy Survey (VIKING; \citealp{Edge2013}), and the VST Kilo-Degree Survey (KiDS; \citealp{deJong2013}). 
These fields are especially valuable due to the extensive supporting data and analysis provided by the Galaxy and Mass Assembly (GAMA) survey \citep{Driver2010}, including highly complete spectroscopic redshifts \citep{Liske2015} and matched-aperture photometry from far-UV to submm \citep{Driver2015}.

In order to unlock the scientific capabilities of these rich data-sets, one of the first challenges is to identify counterparts across the various surveys, a problem of particular difficulty in the submm due to poor angular resolution and the relatively flat redshift distribution of sources in contrast with other wavebands. 
A popular approach to this problem is the likelihood ratio (LR) technique \citep{Richter1975, Sutherland1992, Ciliegi2003}, which was adopted for the
identification of SDSS counterparts in the \hatlas\ science demonstration phase (SDP) by \citet[hereafter \citetalias{Smith2011a}]{Smith2011a}; \textit{Spitzer}-IRAC counterparts in SDP by \citet{Kim2012a}; VIKING counterparts in the DR1 GAMA9 field by \citet{Fleuren2012}; and WISE counterparts in GAMA15 by \citet{Bond2012}.

In this paper we apply the LR technique to \hatlas\ DR1 to find counterparts in SDSS DR7 \citep{Abazajian2009} and DR9 \citep{Ahn2012}, using additional data from UKIDSS-LAS and GAMA. 
For counterparts within the GAMA main survey (primarily $r<19.8$), we take advantage of matched far-UV-to-mid-IR photometry provided by GAMA, combining data from SDSS, GALEX, VIKING and WISE \citep{Driver2015}.
We use the results to determine the multi-wavelength properties and physical nature of 250-\mum-detected sources above $\sim$30~mJy, including magnitude distributions, optical/infrared colours, redshift distribution, and infrared luminosities.
Throughout the paper we quote magnitudes in the AB system unless otherwise stated,
and assume a cosmology with $\Omega_\Lambda=0.73$, $\Omega_M=0.27$ and $H_0=71$\,km\,s$^{-1}$Mpc$^{-1}$.

\section{Data}
\label{sec:data}

\subsection{Submillimetre catalogues}
\label{sec:submm}

We use the \hatlas\ DR1 source catalogues described in \citetalias{Valiante2014}, which are extracted from the 250-\mum\ maps using MADX (Maddox in preparation) 
with a matched-filter technique \citep{Chapin2011} to minimize instrumental and confusion noise.
The source catalogues used for matching contain 37612, 36906 and 39472 sources detected with 250-\mum\ signal-to-noise ratio (SNR) $\geq 4$ in the GAMA9, GAMA12 and GAMA15 fields respectively. The $4\sigma$ detection limit is 24\,mJy for a point source in the deepest regions of the maps (where tiles overlap), or 29\,mJy for a point source in the non-overlapping regions (the average 4$\sigma$ point-source detection has $S_{250}=27.8$\,mJy).
The catalogues also contain a small fraction of sources selected at 350 or 500\mum, which have 250\mum\ SNR~$<4$. We exclude these from this analysis because their red submm colours indicate a high redshift, and any identification with an SDSS source is likely to be false, resulting from either chance alignment or lensing \citep{Negrello2010, Pearson2012}.

For the LR calculations we treat all 250-\mum\ detections as point sources (see Section~\ref{sec:offsets}) but for the discussion of multi-wavelength properties in Section~\ref{sec:multiwlresults} we use the best flux measurement for each source in each band, as described in \citetalias{Valiante2014}.

\subsection{Optical catalogues and classifications}
\label{sec:optical}

\subsubsection{Sample selection}
\label{sec:sample}

\begin{table}
\caption{Statistics of SDSS candidates within the SPIRE masks of the three fields}
\begin{center}
\begin{tabular}{l r r r}
\hline
Sample & GAMA9 & GAMA12 & GAMA15 \\
\hline
SPIRE SNR$_{250}\geq4$ & 37612 & 36906 & 39479 \\
Mask area (deg$^2$) & 53.42 & 53.47 & 54.53\\
SDSS $r_\text{model}<22.4$ & 1127518 & 976822 & 1105073 \\
After cleaning & 1126510 & 975630 & 1103891\\
Stars & 540739 & 365708 & 488648 \\
Galaxies & 754598 & 603327 & 603578 \\
QSOs (spectroscopic) & 3545 & 2921 & 5769 \\
QSOs (photometric) & 7628 & 3674 & 5896 \\
Spectroscopic redshifts & 66368 & 72868 & 81464 \\
\hatlas\ $z_p$ & 1107539 & 963395 & 1092258 \\ 
\hline
\end{tabular}
\label{tab:stats}
\end{center}
\end{table}

We opt to search for counterparts in SDSS because this is currently the deepest comprehensive survey of the full DR1 sky area.
We use all primary objects in SDSS DR7 with $r_\text{model}<22.4$ (see Table~\ref{tab:stats} for numbers). 
We use DR7 because, unlike later releases, it contains size information that is used for identifying \hatlas\ sources requiring extended flux measurements.
We found that 163 and 433 additional objects were present in DR9 (in GAMA9 and GAMA12 respectively), close to bright stars that were masked in DR7. These sources were added to our optical catalogue to maximise the completeness.
We cleaned the catalogue by removing a total of 3749 spurious objects, which are generally either galaxies which have been deblended into multiple primary objects (for example where a spiral arm or bright region has been identified as a separate object from the galaxy) or stars with diffraction spikes which are detected as multiple objects. These cases were identified by visually inspecting all SDSS objects with $r<19$ that had deblend flags and were within 10~arcsec of a SPIRE source; in each case only the brightest central object was retained in the cleaned catalogue (see also \citetalias{Smith2011a}). 
The total of 3749 also includes spurious objects identified by performing a self-match on SDSS coordinates with a search radius of 1 arcsec, since these close pairs are always the result of incorrect deblending in the SDSS catalogue rather than genuine pairs of separate sources.

\subsubsection{Redshifts}
\label{sec:z}

Spectroscopic redshifts ($z_s$) were obtained from the GAMA~II redshift catalogue (\emph{SpecCat v27}), which is close to 100 per cent complete for $r<19.8$ \citep{Liske2015}. 
In addition to GAMA redshifts from the AAO and LT,  we include redshifts collected
from SDSS DR7 and DR10 \citep[both galaxy and QSO targets;][]{Ahn2013}; 
WiggleZ \citep{Drinkwater2010};
2SLAQ-LRG \citep{Cannon2006};
2SLAQ-QSO \citep{Croom2009}; 
6dFGS \citep{Jones2009};
MGC \citep{Driver2005};
2QZ \citep{Croom2004};
2dFGRS \citep{Colless2001};
UZC \citep{Falco1999};
and NED.\footnote{The NASA/IPAC Extragalactic Database, \url{http://ned.ipac.caltech.edu/}, is operated by the Jet Propulsion Laboratory, California Institute of Technology, under contract with the National Aeronautics and Space Administration.}
The GAMA redshifts cover two samples, the main sample and the filler sample.
The main sample is based on SDSS galaxy selection to $r_\text{petro} < 19.8$ primarily. 
In addition to these, \hatlas-selected galaxies were added as filler targets from February 2011. 
Filler targets were selected for \hatlas\ sources with reliable optical
counterparts (reliability $R>0.8$), that were part of the GAMA input catalogue 
(selected from SDSS with $r_\text{petro} < 20.0$ or $r_\text{model} < 20.6$). 
In addition, the same masking and star-galaxy separation were applied to select the filler sample 
as per the GAMA main sample \citep{Baldry2010}. 
Redshifts were considered reliable if they had quality $nQ\geq3$.
The main sample is highly complete with over 98 per cent meeting this criterion, while
86 per cent of the H-ATLAS fillers were observed spectroscopically with 63 per cent having $nQ\geq3$. 
Completeness at $r\gtrsim20$ is lower, although the SDSS DR10 and WiggleZ surveys provide a considerable contribution here (see also Section~\ref{sec:zdist}).
We discarded GAMA redshifts with low quality ($nQ<3$) and SDSS redshifts with any spectroscopic flags set. For objects with redshifts in multiple surveys we favoured the one with the highest quality flag, and where this was not possible we selected in order of preference (i) GAMA, (ii) SDSS DR10, (iii) WiggleZ, (iv) other surveys.
The distribution and origins of redshifts in our final catalogue are described in more detail in Section~\ref{sec:zdist}. 

\begin{figure*}
\begin{center}
\includegraphics[width=\textwidth,clip,trim=18mm 3mm 18mm 3mm]{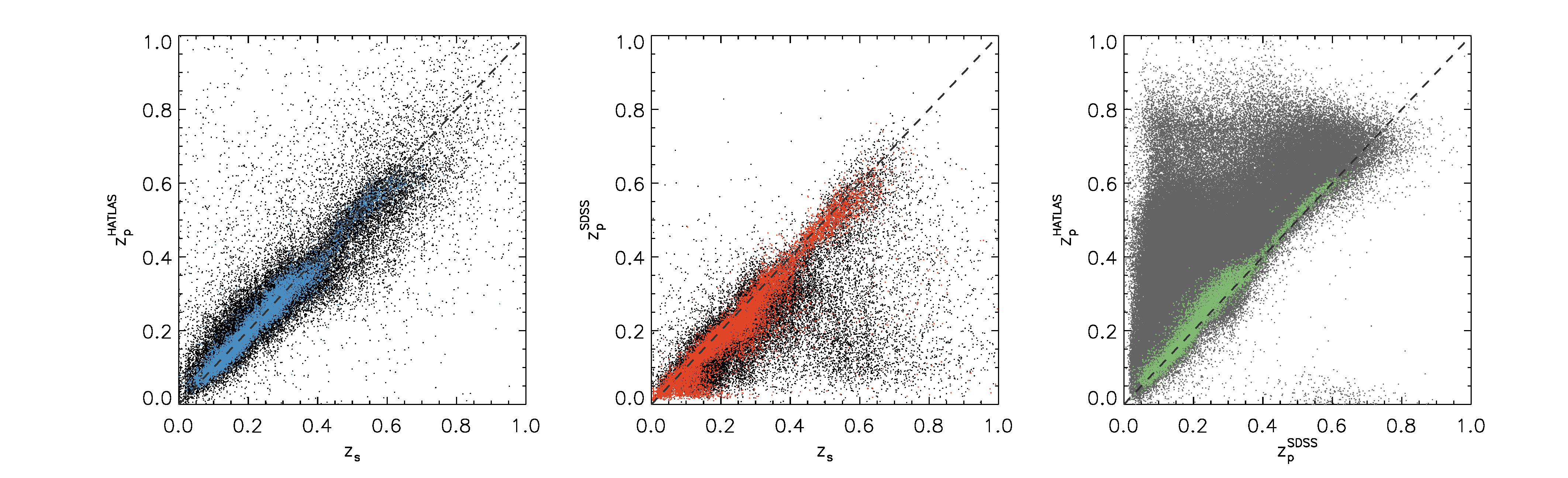}
\includegraphics[width=\textwidth,clip,trim=18mm 3mm 18mm 0mm]{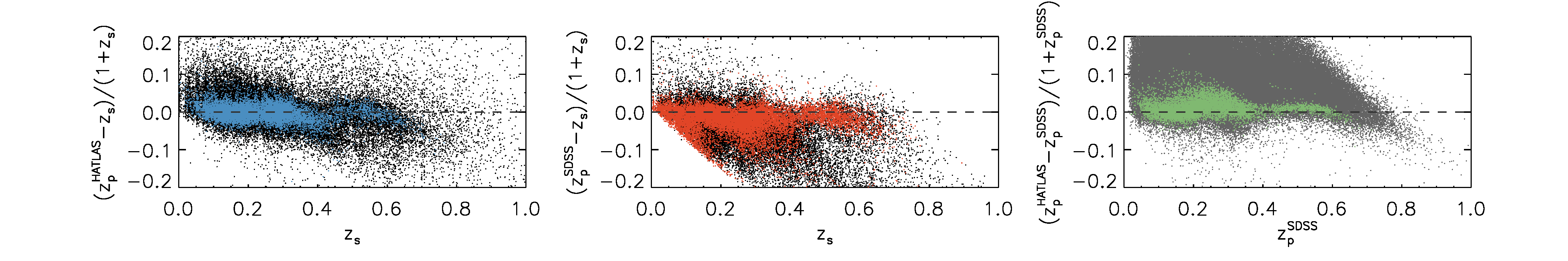}
\caption{Left: comparison of the \hatlas\ photometric redshifts ($z_p$ from $ugrizYJHK$) and available spectroscopic redshifts ($z_s$): black points show the $z_p$ of all galaxies with a $z_s$, and blue show the subset with measurement errors $\Delta z_p/(1+z_p)<0.02$. Middle: comparison of the SDSS photometric redshifts (from $ugriz$) and available spectroscopic redshifts: black points show the $z_p$ of all galaxies with a $z_s$ and red show the subset with measurement errors $\Delta z_p/(1+z_p)<0.02$. Right: direct comparison of the \hatlas\ and SDSS photometric redshifts: grey points show all data with both $z_p$ measurement errors $\Delta z_p/(1+z_p)<0.1$, and green show those with both $z_p$ measurement errors $\Delta z_p/(1+z_p)<0.02$. The lower panels show the respective fractional deviations in $z_p$. To reduce crowding we plot only one quarter of available data in each panel.}
\label{fig:z-z}
\end{center}
\end{figure*}

The optical catalogues also contain photometric redshifts measured from the SDSS $ugriz$ and UKIDSS $YJHK$ photometry for all candidates in our optical catalogue, although these are not used for the LR analysis. 
The photometric redshifts are described in detail in \citetalias{Smith2011a}. They were estimated by empirical regression using the neural network technique of \textsc{annz} \citep{Collister2004}, by constructing a training set of spectroscopic redshifts from GAMA~I \citep{Driver2010}, SDSS DR7, 2SLAQ, AEGIS \citep{Davis2007} and zCOSMOS \citep{Lilly2007}, which covers magnitudes $r<23$ (with $>$1000 galaxies per unit magnitude) and redshifts $z<1$ (with $>$1000 in each $\Delta z$=0.1 bin). 
In Fig.~\ref{fig:z-z} we analyse the accuracy of these photometric redshifts, and compare them against those from the SDSS (DR7/DR9) \emph{Photoz} table\footnote{See \url{http://www.sdss.org/dr7/algorithms/photo-z.html}}. 
The \textsc{annz} redshifts show less bias and smaller scatter than the SDSS ones, showing the benefit of including near-IR photometry from UKIDSS.

\subsubsection{Separating stars, galaxies and quasars}
\label{sec:sgsep}

\begin{figure}
\begin{center}
\includegraphics[width=0.48\textwidth,clip,trim=3mm 3mm 3mm 3mm]{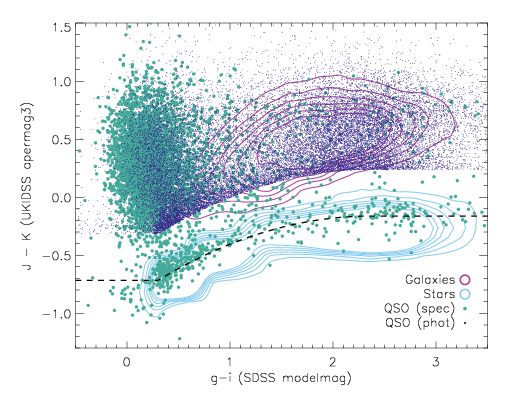}
\caption{Star-galaxy separation using SDSS and UKIDSS colours. The dashed line given by \eqnref{eqn:f_jk} describes the sequence of stars. 
Galaxies (magenta contours) are separated from stars (blue contours) using the criteria described in Section~\ref{sec:sgsep}, based on redshift and size information as well as the colours. 
Quasars are identified as unresolved objects with spectroscopic redshifts $>0.001$ (turqoise circles), and this sample is supplemented with candidate quasars (blue points) which are unresolved objects with non-star-like colours but no secure spectroscopic redshift available.
}
\label{fig:sgsep}
\end{center}
\end{figure}

Star-galaxy separation was carried out using similar constraints to \citetalias{Smith2011a} and the GAMA input catalogue \citep{Baldry2010}. Specifically, galaxies were defined as objects satisfying any of the following constraints, indicating that they are either extended or have colours inconsistent with stars:
\begin{enumerate}
\item $\Delta_{sg} > 0.25$ and not $z_s <0.001$
\item $\Delta_{sg} > 0.05$ and $\Delta_{sgjk} > 0.40$ and not $z_s <0.001$
\item $\Delta_{sg} > f_{sg}$ and not $z_s <0.001$
\item GAMA Kron ellipse defined and not $z_s <0.001$
\end{enumerate}
The $\Delta_{sg}$ parameter quantifies the fraction of extended flux in the $r$-band,
\begin{equation}
\Delta_{sg} = r_\text{psf} - r_\text{model},
\end{equation}
while the $\Delta_{sgjk}$ parameter quantifies the near-IR colour excess,
\begin{equation}
\Delta_{sgjk} = J_{c} - K_{c} - f_{jk},
\label{eqn:delta_sgjk}
\end{equation}
in relation to the colour sequence of stars,
\begin{align}
\nonumber f_{jk} =&\\
\nonumber &-0.7172  \qquad \qquad  \qquad [g_{c}-i_{c}<0.3] \\
\nonumber &-0.89+0.615(g_c-i_c)-0.13(g_c-i_c)^2 \\  
\nonumber & \qquad  \qquad  \qquad \qquad  \qquad[0.3<g_{c}-i_{c}<2.3] \\
&-0.1632  \qquad \qquad  \qquad [g_{c}-i_{c}>2.3].& 
\label{eqn:f_jk}
\end{align}
Finally, $f_{sg}$ provides a stricter threshold on the extended flux fraction at fainter magnitudes where low-surface-brightness emission might otherwise be missed:
\begin{align}
\nonumber f_{sg} =& \\
\nonumber &0.25  \qquad  \qquad \qquad    \qquad     [r_\text{model} < 19]\\
\nonumber &0.25-\dfrac{(r_\text{model}-19)}{15}    \ \ \, [19<r_\text{model}<20.5]\\
&0.15  \qquad  \qquad \qquad  \qquad   [r_\text{model}>20.5]. 
\end{align}
The colour constraints are illustrated in Fig.~\ref{fig:sgsep}.
We used model magnitudes in $g$ and $i$ from SDSS and 2-arcsec aperture magnitudes ({\sc apermag3}) in $J$ and $K$ from UKIDSS-LAS;\footnote{Note that GAMA aperture photometry is only available for $r<19.8$ galaxies.} the subscript $c$ denotes correction for Galactic extinction following \citet{Baldry2010}, using the extinction maps from \citet{Schlegel1998}.

Quasars (QSOs) were identified as SDSS objects which do not satisfy any of the criteria (i--iv) above (i.e. unresolved in the $r$-band and with non-stellar colours), and which have secure spectroscopic redshifts $z>0.001$. 
We also classified as quasars all objects which had been classified as quasars by their SDSS spectra, defined by the criteria {\sc class}=`QSO' and {\sc zwarning}=0 in the \emph{SpecObj} table of SDSS DR12.
In addition to these `spectroscopic' QSOs, we identified photometric QSO candidates as objects which do not satisfy (i--iv), have no spectroscopic redshift, but have $\Delta_{sgjk}>0.40$ (i.e. non-stellar colour).
The numerical results of each classification are shown in Table~\ref{tab:stats}.

\section{Likelihood ratio analysis}
\label{sec:methods}

The identification of counterparts (`IDs') to submm sources can be tackled in a statistical way using the LR technique to assign a probability (`reliability') to each potential match, and thus distinguish robust counterparts from chance alignments with background sources. 
The LR method relies on knowledge of the intrinsic positional uncertainty of the sources as well as the magnitude distributions of true counterparts and background sources.
When the sample size is sufficiently large it is preferable to divide the matching catalogue into different source classifications to measure these statistics for each class of object (\citealp{Sutherland1992}; \citealp[\citetalias{Smith2011a};][]{Chapin2011,Fleuren2012}). 
We therefore calculate magnitude statistics separately for stars and for extragalactic objects (galaxies and QSOs, both spectroscopic and photometric). We assume that positional errors are independent of the source classification, but we take into account the potential bias for redder submm sources that was highlighted by \citet{Bourne2014}, as we shall describe in Section~\ref{sec:offsets}.

The LR of each counterpart is defined as a function of $r$-band magnitude ($m$) and radial separation ($r$), as
\begin{equation}
 L = \dfrac{q(m,c) f(r)}{n(m,c)},
 \label{eqn:lr}
\end{equation}
where $q(m,c)$ represents the magnitude distribution of counterparts in class $c$ (i.e. stars or galaxies), $n(m,c)$ represents the background magnitude distribution in class $c$, and $f(r)$ represents the positional offset distribution due to positional errors in both catalogues (in this case we can assume that the optical positional errors are negligible compared with those of SPIRE). We will describe the measurement of each of these quantities in the following subsections.

\subsection{Magnitude distributions}
\label{sec:q}

The normalized magnitude distribution $n(m)$ of SDSS objects (either stars or galaxies) describes the probability density that a given optical source has magnitude $m$. 
We must also measure the distribution $q(m)$, which describes the probability density that the true optical counterpart to a given SPIRE source has magnitude $m$. 
This is given by normalizing the magnitude histogram of all true counterparts, $n_\text{real}(m)$, and scaling by the overall probability $Q_0$ that the counterpart is detected in SDSS:
\begin{equation}
 q(m) = Q_0\, \dfrac{n_\text{real}(m)}{\sum_m{n_\text{real}(m)}}.
\end{equation}
The top panel of Fig.~\ref{fig:qm} illustrates the magnitude histograms of various samples used to estimate this.
The magnitude distribution of true counterparts, $n_\text{real}(m)$, was measured by taking the magnitude histogram of all optical sources within 10 arcsec of any SPIRE position [solid histograms; $n_\text{total}(m)$], and subtracting the background magnitude histogram of all optical sources scaled to the same search area [dotted histograms; $n_\text{bkgrd}(m)$]. 
Each of these histograms is measured separately for stars and extragalactic sources as shown in Fig.~\ref{fig:qm}, since the optical magnitudes of submm-detected stars and of submm-detected galaxies are very different (as shown in the middle panel).

\subsubsection{The normalization $Q_0$}
\label{sec:q0}
The normalization of the probability distribution $q(m)$ is $Q_0$, which is equal to the fraction of all true counterparts which are above the SDSS magnitude limit. This was calculated separately for stars and for extragalactic objects following the method of \citet{Fleuren2012}. 
This method involves measuring $1-Q_0$ by counting blanks (sources without counterparts) as a function of search radius, thus removing the possibility of bias due to clustering or multiple counterparts (which would boost a direct measurement of $Q_0$). The number of blanks as a function of search radius is plotted in Fig.~\ref{fig:q0}, and we fit this with the model
\begin{equation}
B(r)=1-Q_0 F(r)
\label{eqn:q0model}
\end{equation}
\citep{Fleuren2012}, where $F(r)$ is the distribution of radial offsets to counterparts, given by
\begin{equation}
 F(r) = \int_0^r 2\pi r^\prime f(r^\prime)dr^\prime = 1-\exp(-r^2/2\sigma^2),
 \label{eqn:blanksmodel}
\end{equation}
and $f(r)$ is the Gaussian positional error function defined by \eqnref{eqn:f_r} in Section~\ref{sec:offsets}.
We thus measure $Q_0=0.519\pm0.001$ for QSOs and galaxies, while for stars we measure $Q_0=0.020\pm0.002$.
In comparison, \citetalias{Smith2011a} measured $Q_0=0.583$ for galaxies and $Q_0=0.010$ for unresolved sources (both stars and QSOs), using about 10 per cent of the data used here. The decrease in $Q_0$ for galaxies between the SDP \citepalias{Smith2011a} and the current DR1 data 
can be explained by the increased depth of the catalogues (28mJy on average compared with 32mJy in SDP, leading to a higher fraction of high-redshift submm sources without SDSS counterparts), combined with the fact that our method accounts for the bias caused by clustering and multiple sources (while \citetalias{Smith2011a} did not), which is likely to affect the $Q_0$ for galaxies. 

We can show this by instead calibrating $Q_0$ from the normalization of the magnitude histograms (as in \citetalias{Smith2011a}), giving $Q_0=0.616$ for galaxies, suggesting that the level of bias from clustering and multiple sources is significant. A further independent measurement can be obtained from the normalization of the positional offset histograms (see Section~\ref{sec:offsets}). 
This measurement is $Q_0=0.548\pm0.009$ for galaxies, which accounts for clustering via a cross-correlation term but can still be boosted by multiple counterparts, unlike the blanks method. Comparing the various methods, we might suggest that $Q_0$ can be boosted by 0.03 due to multiplicity, and by a further 0.07 due to clustering, although these estimates are rough and dependent on the models assumed (for example the power-law correlation function assumed in Section~\ref{sec:offsets}).
We will discuss the sensitivity of the results to the value of $Q_0$ in Section~\ref{sec:lr-q0}.

Field-to-field variance is also likely to be significant between the 16~deg$^2$ SDP field and the 161~deg$^2$ in DR1, and this probably accounts for the different measurements for stars. The three GAMA fields sample quite different Galactic latitudes and hence sight-lines through the disk and halo of the Milky Way: GAMA9 is at $b=+28$, GAMA12 is at $b=+60$ and GAMA15 is at $b=+54$.
The standard deviation in $Q_0$ from the blanks method between the three DR1 fields (each $\approx$53~deg$^2$) is 0.011 for galaxies and 0.011 for stars. 

\subsubsection{Application to the likelihood ratios}
\label{sec:qm}
\begin{figure}
\begin{center}
\includegraphics[width=0.48\textwidth,clip,trim=6mm 3mm 6mm 3mm]{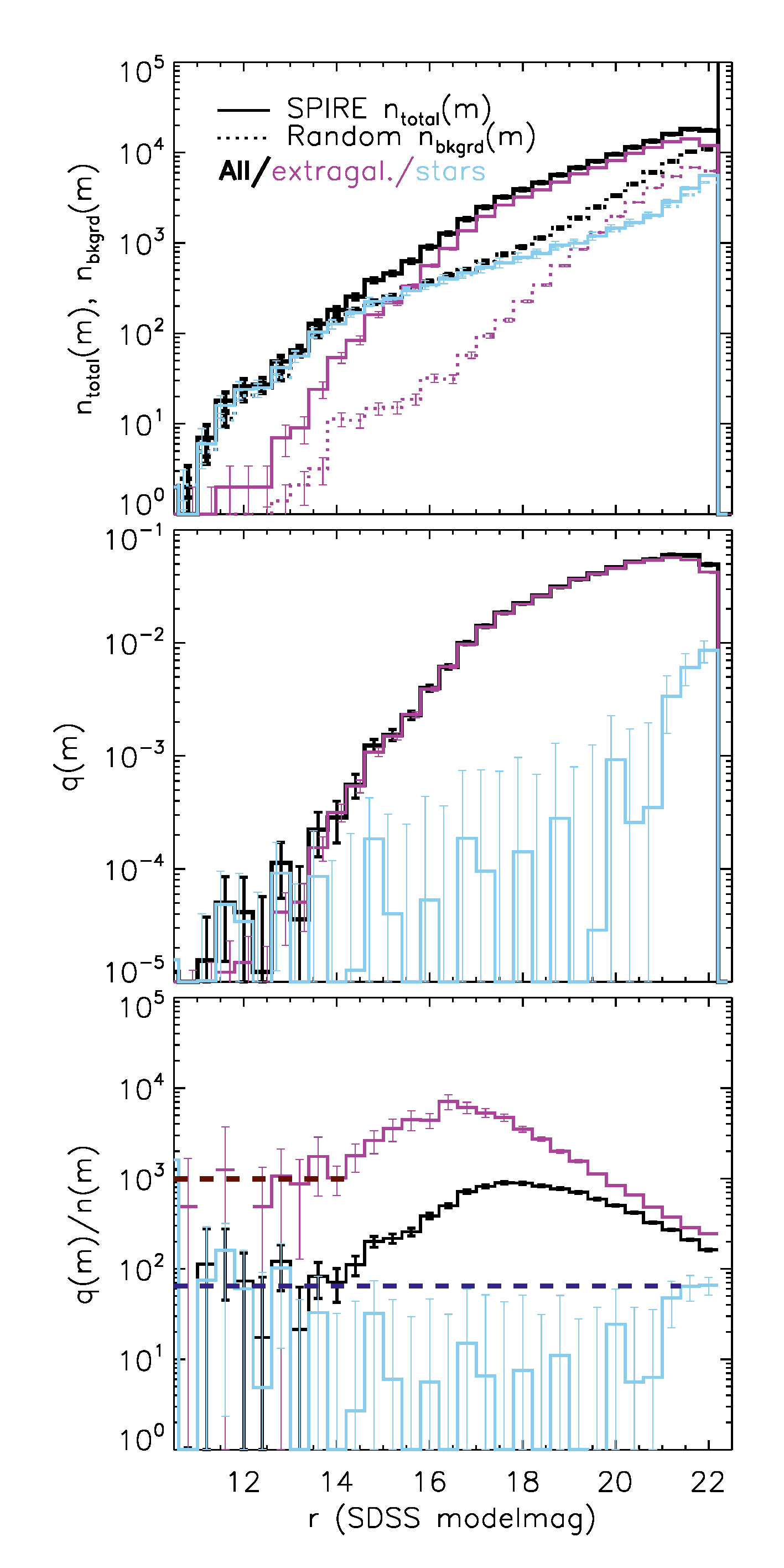}
\caption{The method of measuring the $r$-band magnitude distribution of true counterparts. Top panel: magnitude distribution of SDSS objects within 10~arcsec of all SPIRE positions (solid lines) divided into galaxies and quasars (magenta) and stars (blue), and the total (heavy black line). 
The background magnitude distributions of SDSS (normalized to the same search area) are given by the dashed lines (the background distribution of stars closely follows that of SPIRE centres).
Middle panel: the magnitude distribution of true counterparts $q(m$) is given by the difference between SPIRE and random distributions above, normalized to $Q_0$.
Bottom panel: the ratio of the magnitude distribution of true counterparts to that of background objects is used in the calculation of LR in \eqnref{eqn:lr}. At magnitudes $m<14.0$ (galaxies) and $m<21.5$ (stars)  
the value of $q(m)/n(m)$ is fixed at the average within this range, as shown by the horizontal dashed lines.
}
\label{fig:qm}
\end{center}
\end{figure}

\begin{figure}
\begin{center}
\includegraphics[width=0.48\textwidth,clip,trim=6mm 3mm 3mm 3mm]{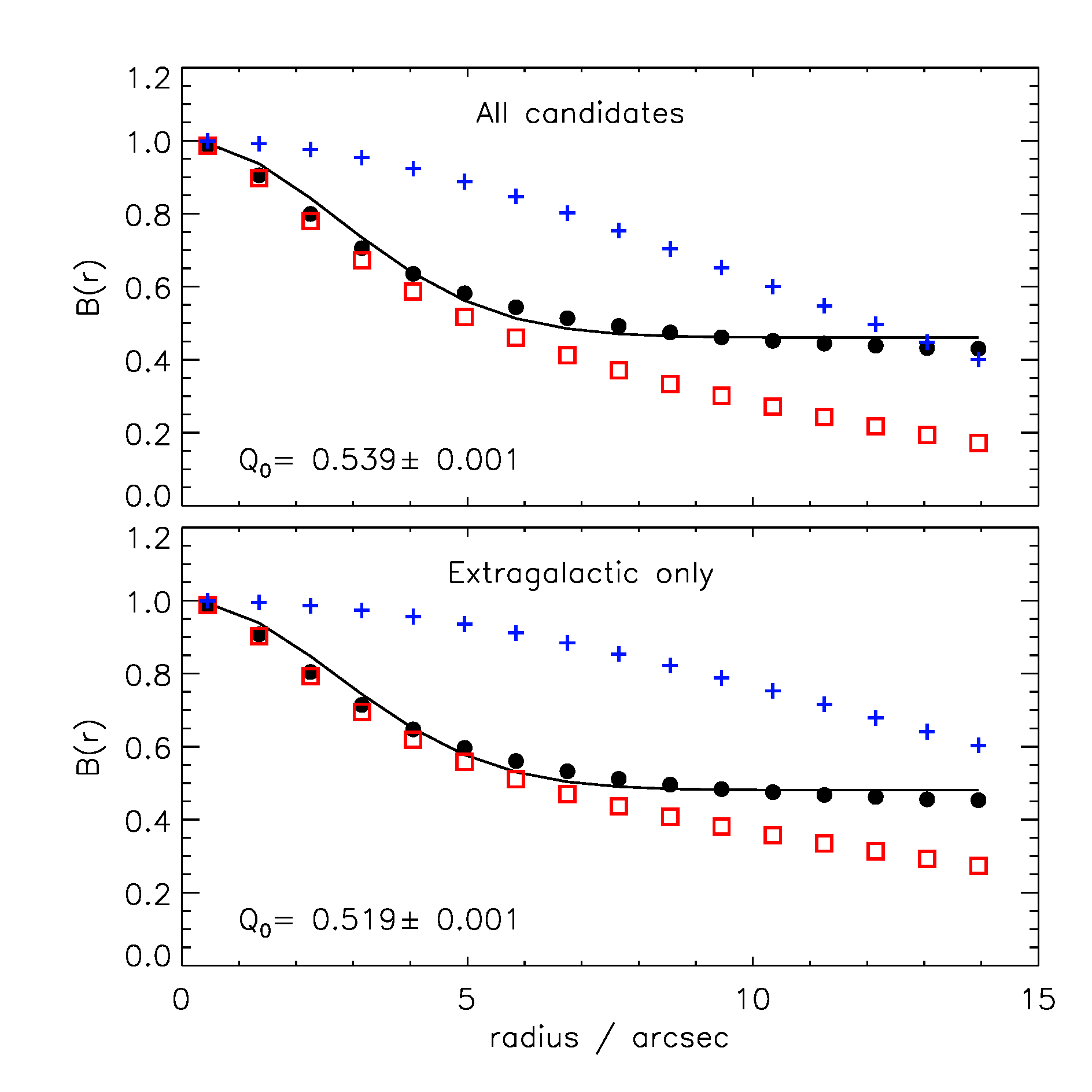}
\caption{The method of measuring the fraction of SPIRE sources without counterparts ($1-Q_0$) by counting `blanks' (objects with no candidate within the search radius) as a function of the search radius. Red squares show the blank counts centred on real SPIRE positions, blue crosses show the blank counts around random positions, and
black circles show the ratio. The black line is the best fit to the black circles using the model described by \eqnref{eqn:q0model}.
Top panel: all optical candidates; bottom panel: galaxies and quasars only.}
\label{fig:q0}
\end{center}
\end{figure}

We take the difference between the SPIRE-centred and background magnitude histograms, $n_\text{real}(m)$, and normalize this to $Q_0$ to give $q(m)$, as shown in the middle panel of Fig.~\ref{fig:qm}. 
In calculating the LR for extragalactic candidates, we use the measured $q(m)/n(m)$ distribution for $r$-band magnitudes $m>14.0$, but at brighter magnitudes the distribution is not well sampled and we assume a constant $q(m)/n(m)$ equal to the average of $q(m)/n(m)$ at $m<14.0$. 
The same method is used for stellar candidates, except that the threshold must be set at $m=21.5$ because there are too few to measure the magnitude distributions at $m<21.5$.
The measured distributions of $q(m)/n(m)$ are shown by the histograms on the lower panel of Fig.~\ref{fig:qm}, while the constant values adopted for bright magnitudes are indicated by the horizontal dashed lines.
In this respect our method differs from \citetalias{Smith2011a}, who assumed a constant $q(m)$ for stars [as opposed to constant $q(m)/n(m)$], which would lead to a higher LR for brighter stars compared with fainter ones since $n(m)$ rises towards fainter magnitudes. Our assumption of flat $q(m)/n(m)$ at $m<21.5$ instead leads stars to have an LR independent of magnitude. 
Our motivations for this choice are (i) that the use of a constant $q(m)$ [hence non-constant $q(m)/n(m)$] at $m<21.5$ would lead to a discontinuity in the lower panel of Fig.~\ref{fig:qm} where the decreasing $q(m)/n(m)$ at increasing $m$ meets the measurements at $m>21.5$ which are relatively high; and (ii) that we do not necessarily expect any correspondence between optical magnitude and the detectability of a star at 250\mum.

This last point deserves some explanation. The submm emission associated with stars is likely to come from debris discs or dust in outflows and so may not be directly related to the photospheric optical luminosity. We might still expect a correlation between optical and submm fluxes because the dust emission in these systems is simply reprocessed starlight, and also because the flux in both wavebands depends on distance. 
However, there is a large amount of scatter in the masses and temperatures of debris discs around stars of a given spectral type \citep{Hillenbrand2008, Carpenter2009}, and \hatlas\ will detect only the brightest of these \citep{Thompson2010}. The statistics of \hatlas-detected stars will therefore be highly stochastic and any correlation between submm detectability and optical magnitude is likely to be broken.
The effect of this assumption on LR statistics is discussed in Section~\ref{sec:stellarids}.

Finally, in Section~\ref{sec:checks} we visually check the positions of bright stars to ensure none are missing from the ID catalogue. We find that the few bright stars detected at 250\mum\ are successfully identified by the LR procedure, and none of them are missing from the SDSS catalogue. This indicates that the measured $q(m)$ for bright stars is not under-estimated due to incompleteness of SDSS at bright magnitudes.

\subsection{Positional offset distributions}
\label{sec:offsets}

The final ingredient for measuring the LR of each potential counterpart is the probability that a real counterpart is found at the measured radial separation $r$ from the SPIRE source.
For the purposes of assigning likelihood ratios to counterparts we make the simplifying assumption that all \hatlas\ sources are point-like in the 250\,\mum\ maps from which they were extracted, with a beam full-width at half-maximum of 18.1~arcsec. This assumption is justified by the fact that a maximum of two per cent of the sources are resolved,\footnote{This is the fraction (2572/113997) of SPIRE 250-\mum\ sources which have reliable optical counterparts with $r<19$ and ISOA\_r$>10$~arcsec, although only 710 of them are treated as resolved for photometry purposes in \citetalias{Valiante2014}, while the number of sources that are significantly extended beyond a single beam is much smaller.} 
and that these will be galaxies at $z<0.05$ with bright, unambiguous SDSS counterparts. 

The distribution of angular separations ($\Delta RA,\Delta Dec$) of true counterparts follows the probability distribution function of random positional errors, which can be described by a Gaussian in two dimensions:
\begin{equation} 
f(x,y) = 
 \dfrac{1}{{2\pi}\sigma_{x} \sigma_{y}} \exp{\left( \dfrac{-x^2}{2\sigma_{x}^2}\right)} \exp{\left( \dfrac{-y^2}{2\sigma_{y}^2}\right)}.
 \label{eqn:f_xy}
\end{equation}
These positional errors are dominated by the errors in SPIRE source extraction (the SDSS positional error is negligible in comparison for all but the most extended galaxies), and since the SPIRE PSF has good annular symmetry we can assume that the positional errors follow a radial form with Gaussian width $\sigmapos=\sigma_x=\sigma_y$, i.e.
\begin{equation} 
 f(r) = \dfrac{1}{{2\pi}\sigmapos^2} \exp{\left(\dfrac{-r^2}{2\sigmapos^2}\right)}.
 \label{eqn:f_r}
\end{equation}
In \eqnref{eqn:lr}, this is normalized such that $\int f(r) dr=1$.
The width \sigmapos\ can be measured by examining the offset distribution of all potential counterparts and modelling the contribution from real counterparts and other correlated sources. Following the methods of \citetalias{Smith2011a} and \citet{Bourne2014},
we counted all SPIRE--SDSS pairs up to a 50~arcsec separation in RA and Dec, to create a two-dimensional histogram of the offsets to SDSS sources around all SPIRE sources. 
This histogram contains at least three contributing populations: 
\begin{enumerate}
\item The random background of chance alignments, which is effectively constant across the histogram since the probability of chance alignment is equal in any given sightline
\item True SDSS counterparts, which exist for a fraction $Q_0$ of all SPIRE sources, and whose distribution follows the positional error function $f(r)$ 
\item Nearby SDSS objects that are physically correlated with the SPIRE source (due to galaxy clustering), but are not direct IDs, whose distribution is given by the cross-correlation between the SPIRE and SDSS samples, governed by a power law $w(r)$ convolved with $f(r)$ 
\end{enumerate}
These components are described by the equation: 
\begin{equation}
n(\Delta RA, \Delta Dec) = n_0+Q_0 f(r)+w(r)*f(r).
\label{eqn:2dhist}
\end{equation}
Additionally, it was shown by \citet{Bourne2014} that these distributions are likely to contain significant contributions from foreground structure in the line of sight to the SPIRE source, which is not physically associated but is  
lensing the source [so that the alignments are not purely random, and are distinct from item (i) above]. \citet{Bourne2014} showed that measured positional errors, even accounting for the galaxy cross-correlation term, are over-estimated by a factor that increases with both brightness and redness of the SPIRE source, which is most likely due to the increased probability of such sources to be lensed. This bias could lead to increased LR values for SDSS associations to red SPIRE sources which are more likely to be lenses, or to be within lensing large-scale structures, rather than being true counterparts.

We can avoid this bias by measuring the positional errors in the subset of blue SPIRE sources with $S_{250}/S_{350}>2.4$, which were found by \citet{Bourne2014} to be minimally biased by lensing (due to their lower redshifts), and applying these measured positional errors to SPIRE sources of all colours.
In theory, the width of the positional errors, \sigmapos, depends on the full-width at half-maximum (FWHM~$=18.1$\,arcsec) and the SNR of the 250\,\mum\ detection;
\begin{equation}
\sigmapos(\text{SNR}) = 0.6 \dfrac{\text{FWHM}}{\text{SNR}}
\label{eqn:ivison2007}
\end{equation}
\citep[Maddox in preparation]{Ivison2007}, although the real \sigmapos\ will be increased by factors related to the map-making procedure and confusion noise \citep[][]{Hogg2001,Negrello2005,Chapin2011,Bourne2014}.
We therefore measure the empirical SNR dependence from offset histograms of blue SPIRE sources in four bins of SNR (with boundaries at SNR~$=4,5,6,8,12$), as described in the following sections. 

\subsubsection{Measuring the cross-correlation}
\label{sec:x}
\begin{figure*}
\begin{center}
\includegraphics[width=\textwidth,clip,trim=3mm 6mm 25mm 3mm]{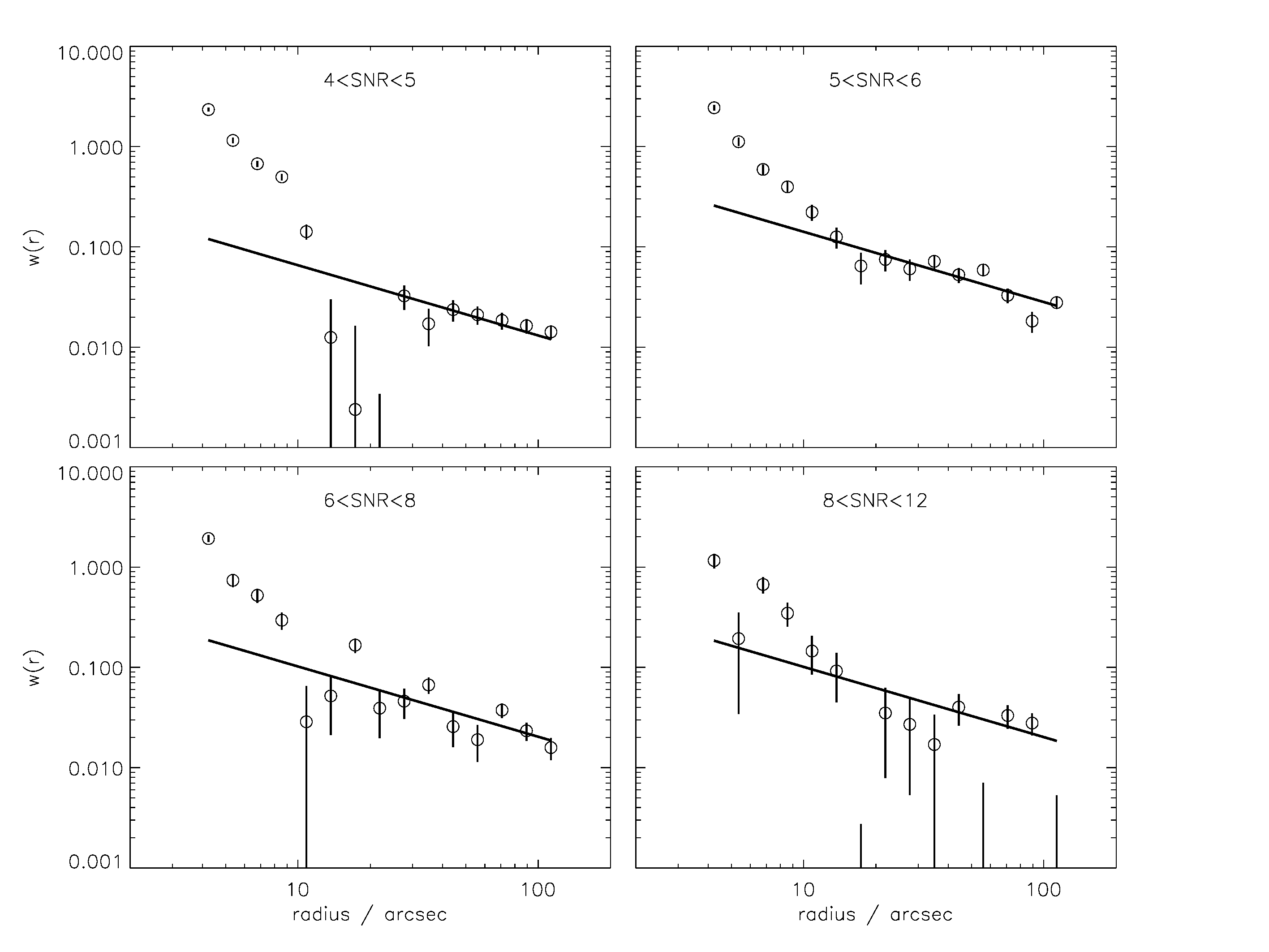}
\caption{Cross correlations between the SPIRE 250-\mum\ positions in bins of SNR and the SDSS positions across the three fields. Error bars represent Poisson counting errors. The power-law fits shown include only the data at $r>10$~arcsec in order to remove any bias from the true counterparts which are at smaller radii. In these fits the slopes are fixed to $-0.7$ and the best-fitting normalization parameters are given in Table~\ref{tab:params}.}
\label{fig:xcorr}
\end{center}
\end{figure*}

We first measure the angular cross-correlation function between (blue) SPIRE and SDSS samples in each SNR bin. 
This is modelled as a power-law
\begin{equation}
w(r)=(r/r_0)^\delta,
\label{eqn:wtheta}
\end{equation}
as a function of radial offset $r$ up to 120~arcsec. We estimate $w(r)$ with a modified \citet{Landy1993} estimator 
to count pairs between the SPIRE and SDSS data ($D_1, D_2$), and random positions ($R$), as a function of radial separation:
\begin{equation}
w(r)= \dfrac{ D_1D_2(r) -D_1R(r) - D_2R(r) + RR(r)}{RR(r)}.
\label{eqn:landy}
\end{equation}
The results are combined across all fields and are shown in Fig.~\ref{fig:xcorr} for each SNR bin. The expected power-law behaviour is seen at large radii, although the apparent steepening at small radii may be due to bias from the offsets to true associations, which would follow the distribution $f(r)$. To avoid this possibility we fit the data at $r>10$~arcsec and find a power-law index consistent with $-0.7\pm0.1$ in each SNR bin \citep[equal to the expected value for galaxies, e.g.][]{Connolly2002}.  We therefore fix the index to this value and obtain best-fitting values for the correlation length as shown in Table~\ref{tab:params}.

\subsubsection{Measuring the positional errors}
\label{sec:sigma}
\begin{figure*}
\begin{center}
\includegraphics[width=\textwidth,clip,trim=3mm 0mm 0mm 3mm]{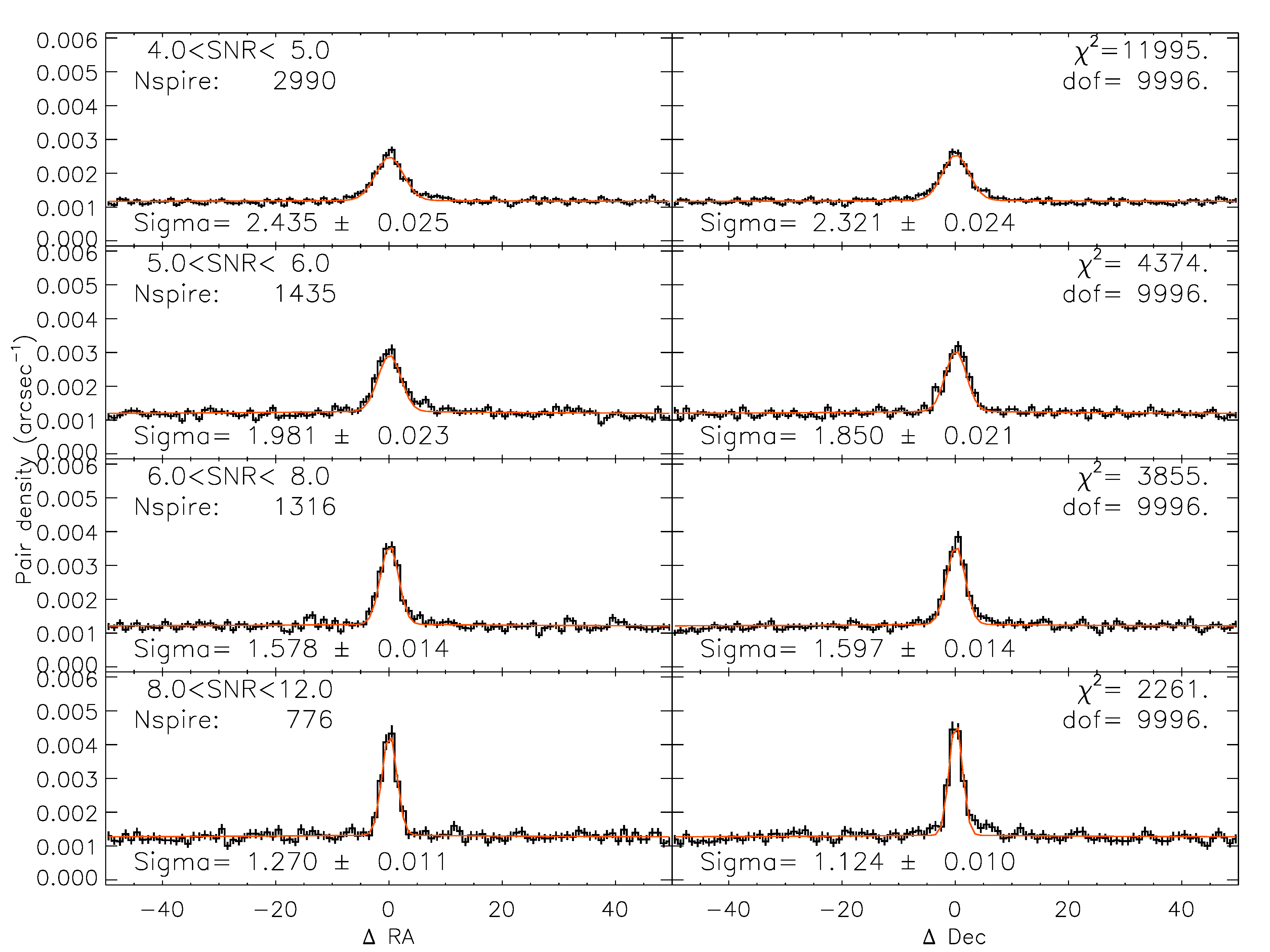}
\caption{The two-dimensional histograms of positional offsets between SPIRE sources in the blue bin ($S_{250}/S_{350}>2.4$), and all SDSS candidates within $\pm50$~arcsec in RA and Dec are shown collapsed along the RA and Dec axes (left and right respectively), divided into bins of SPIRE SNR (from top to bottom). The orange line shows the best-fitting model given by \eqnref{eqn:2dhist} with the widths \sigmapos\ (arcsec) and the $\chi^2$ and degrees of freedom (dof) printed in the respective panels.}
\label{fig:sigmapos_snr_c5}
\end{center}
\end{figure*}

Finally we fit the two-dimensional histograms of $\Delta$RA and $\Delta$Dec separations with the model in \eqnref{eqn:2dhist}, 
fixing $w(r)$ with the power-law parameters described above, and solving for the widths ($\sigma_\text{pos,RA}, \sigma_\text{pos,Dec}$) of the Gaussian $f(r)$, the background level $n_0$ and the normalization $Q_0$. 
We do this in each of the SNR bins, using only blue sources with $S_{250}/S_{350}>2.4$, to avoid the lensing-related biases discussed in \citet{Bourne2014} and in Section~\ref{sec:offsets} above. 
The fitting is performed on the two-dimensional histograms, and in Fig.~\ref{fig:sigmapos_snr_c5} we show the histograms and best-fitting models collapsed along the $\Delta$RA and $\Delta$Dec axes for visual inspection. The fitting results are summarised in Table~\ref{tab:params}. Note that the normalization ($Q_0$) fitted to these histograms differs from the results in Section~\ref{sec:q0} because it depends on both SPIRE colour and SNR, but this normalization can also be boosted by the existence of multiple counterparts.
We will investigate the dependence of our ID results on the measurement of $Q_0$ in Section~\ref{sec:lr-q0}.

\begin{figure*}
\begin{center}
\includegraphics[width=\textwidth,clip,trim=15mm 10mm 0mm 6mm]{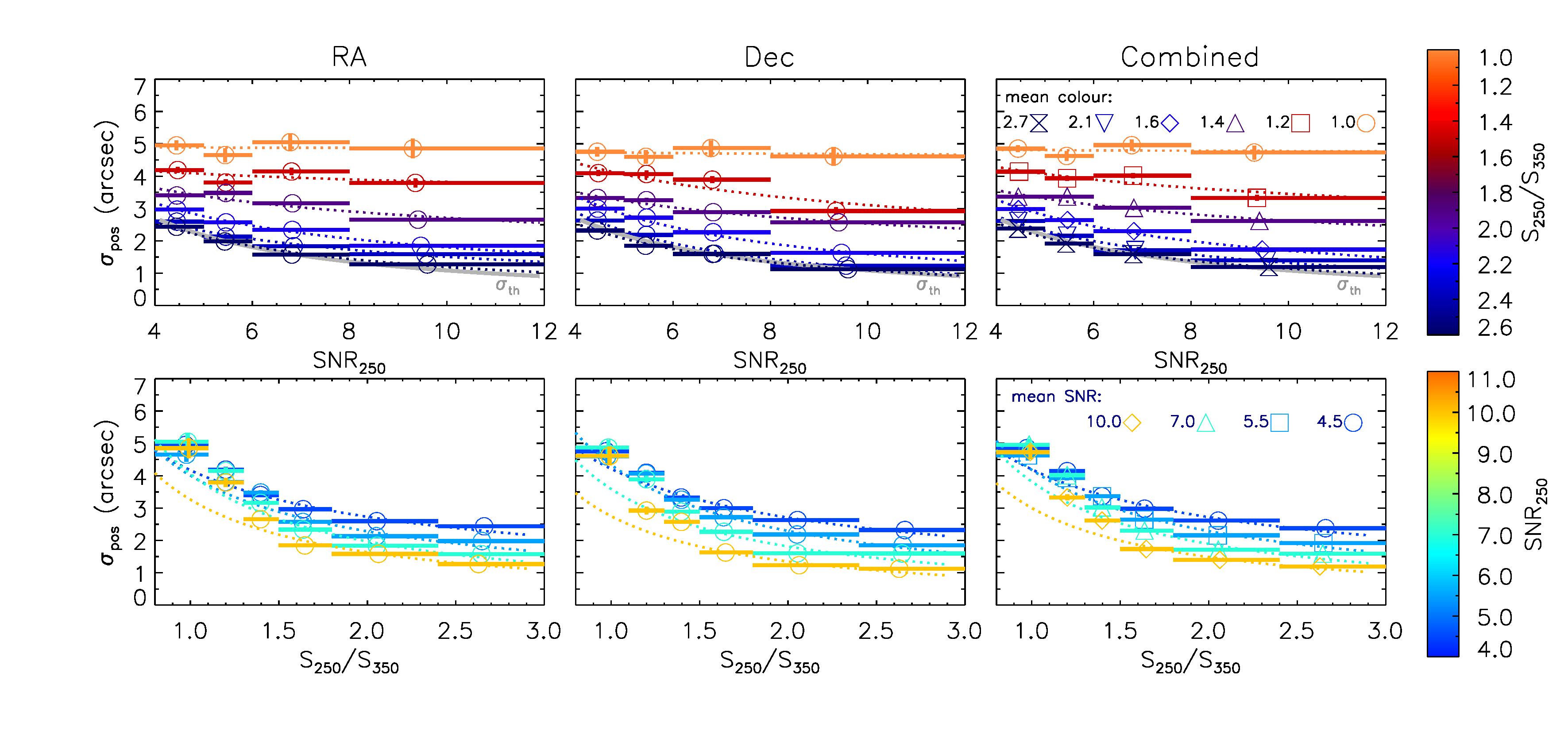}
\caption{Dependence of the positional error on SPIRE SNR and colour. Top panels: the best-fitting \sigmapos\ values (RA, Dec, and the geometric mean thereof) from the two-dimensional offset histograms, as a function of SPIRE SNR, for sources in six bins of 250-/350-\mum\ colour. The bluest bin is the data plotted in Fig.~\ref{fig:sigmapos_snr_c5}. Bottom: the same data plotted as a function of colour in the four SNR bins. The dotted lines show the best-fitting power-law models for \sigmapos\ as a function of SNR in each colour bin (top panels), and as a function of colour in each SNR bin (bottom panels).}
\label{fig:sigmapos_snr_col}
\end{center}
\end{figure*}

\citet{Bourne2014} showed the colour dependence of the width \sigmapos\ by dividing the sample into six bins of colour and four of SNR. We repeat this analysis on the matched-filter catalogues used here and show the best-fitting \sigmapos\ values in each bin in Fig.~\ref{fig:sigmapos_snr_col}. In this analysis the bluest colour bin contains the results shown in Fig.~\ref{fig:sigmapos_snr_c5}. The increasing \sigmapos\ and weakening SNR dependence in redder colour bins demonstrates the bias which is likely due to lensing, and the need to measure the true value of \sigmapos\ using blue SPIRE sources only. 

To describe the SNR dependence of the positional error we fit the best-fitting \sigmapos\ values as a function of mean SNR with the function 
\begin{equation}
\sigmapos(\text{SNR}) = \sigmapos(5) \left[ {\text{SNR}}/{5} \right]^{\alpha}.
\label{eqn:sigmapos_snr}
\end{equation}
The dotted lines in Fig.~\ref{fig:sigmapos_snr_col} show the best-fitting power-law models in each colour bin.
Combining results from the $\Delta$RA and $\Delta$Dec offsets in the bluest bin ($S_{250}/S_{350}>2.4$), we conclude that the positional error for SPIRE sources at SNR~$=5$ is 
$\sigmapos(5)=2.10\pm0.01$~arcsec and that this decreases with SNR raised to the power 
$\alpha=-0.88\pm0.01$.
This empirical SNR dependence is very close to the theoretical inverse relation of \eqnref{eqn:ivison2007}, shown by the the thick grey line labelled $\sigma_{th}$ in Fig.~\ref{fig:sigmapos_snr_col}. The fact that it is slightly shallower could be explained by additional positional errors affecting the brightest sources, such as a non-Gaussian PSF, the 6-arcsec pixel resolution of the 250\mum\ map from which sources were extracted, and the fact that bright sources are not truly point-like at scales of $\lesssim3$~arcsec.
We use these measured parameters in \eqnref{eqn:sigmapos_snr} to compute $f(r)$ for the LR calculations, but we assign a minimum positional error of 1~arcsec since the formula above would give unrealistically small errors for the brightest sources.
We also allow for larger positional errors for more extended galaxies, which may be extended in the submm as well as having a larger uncertainty on the optical position. For galaxies with $r$-band magnitude $m<20.5$ we add (in quadrature) a positional error equal to five per cent of the $r$-band ISOA parameter (isophotal semimajor axis) from SDSS.

\begin{table}
\caption{Best-fitting parameters in modelling SDSS positional offsets to blue SPIRE sources
with $S_{250}/S_{350}>2.4$ in bins of 250\,\mum\ SNR
}
\begin{center}
\begin{tabular}{c c c c c}
\hline
SNR & $N_\text{SPIRE}$ & $r_0/$arcsec$^\text{a}$ &  $\sigmapos/$arcsec$^\text{b,c}$ & $Q_0\,^\text{b}$\\
\hline
4--5  & 2990 & $0.20\pm0.02$   & $2.38\pm0.02$ & $0.792\pm0.008$ \\ 
5--6  & 1435 & $0.61\pm0.05$   & $1.91\pm0.02$ & $0.846\pm0.010$ \\ 
6--8 &  1316 & $0.38\pm0.05$   & $1.59\pm0.01$ & $0.944\pm0.008$ \\
8--12 & 776  & $0.38\pm0.08$   & $1.19\pm0.01$ & $0.985\pm0.008$ \\
\hline
\end{tabular}
\end{center}
Notes: (a) normalization of the cross-correlation function, \eqnref{eqn:wtheta}, with index $\delta=-0.7$;
(b) width $\sigmapos$ and normalization $Q_0$ in \eqnref{eqn:2dhist}; 
(c) circularized values equal to $\sqrt{\sigmapos(\text{RA})\,\sigmapos(\text{Dec})}$.
\label{tab:params}
\end{table}

\subsubsection{Robustness of the positional errors}
The positional error for a 5-$\sigma$ source given above ($2.10\pm0.01$~arcsec) differs from the value of $2.40\pm0.09$~arcsec used in the SDP (\citetalias{Smith2011a}) for the following reasons: 
firstly \citetalias{Smith2011a} measured \sigmapos\ for all sources with SNR$>5$, rather than binning by SNR and colour, which would lead them to measure a larger overall \sigmapos\ due to the bias from red sources as discussed above. Additionally, the use of a matched filter in the DR1 source extraction has improved the positional accuracy compared with the PSF-filtering used in SDP (\citetalias{Valiante2014}), and this reduces the measured \sigmapos. We validated our \sigmapos\ measurement by using the same flux cut as \citetalias{Smith2011a} on catalogues made from PSF-filtered DR1 maps, with no SNR/colour binning, and found a value of $\sigmapos=2.39$~arcsec, consistent with 2.40~arcsec in \citetalias{Smith2011a}. 

It is apparent in Fig.~\ref{fig:xcorr} that there is the potential for degeneracy between the parameters of the cross-correlation (which dominates radial separations on large scales) and those of the positional error (which dominates on small scales). To test what effect the fit to the cross-correlation has on the fit for \sigmapos, we compare the results given by fitting the cross-correlation on different scales. The results given above were obtained from a cross-correlation fit with a slope of $\delta=-0.7$ on scales $r>10$~arcsec. If we fit the cross-correlation to scales of $r>5$~arcsec we obtain a slope of $-1.9$ ($r_0=5.23$~arcsec) and the resulting change in the fit for the positional error with this alternative $w(r)$ is such that \sigmapos\ is reduced by a factor 0.91, and $f(r)$ and $L$ would therefore be boosted by a factor 0.97. We therefore conclude that our results are not strongly affected by the choice of correlation function assumed.

\section{Results}
\subsection{Likelihood ratios and reliabilities}
\label{sec:lr}
\begin{figure}
\begin{center}
\includegraphics[width=0.48\textwidth,clip,trim=3mm 3mm 3mm 3mm]{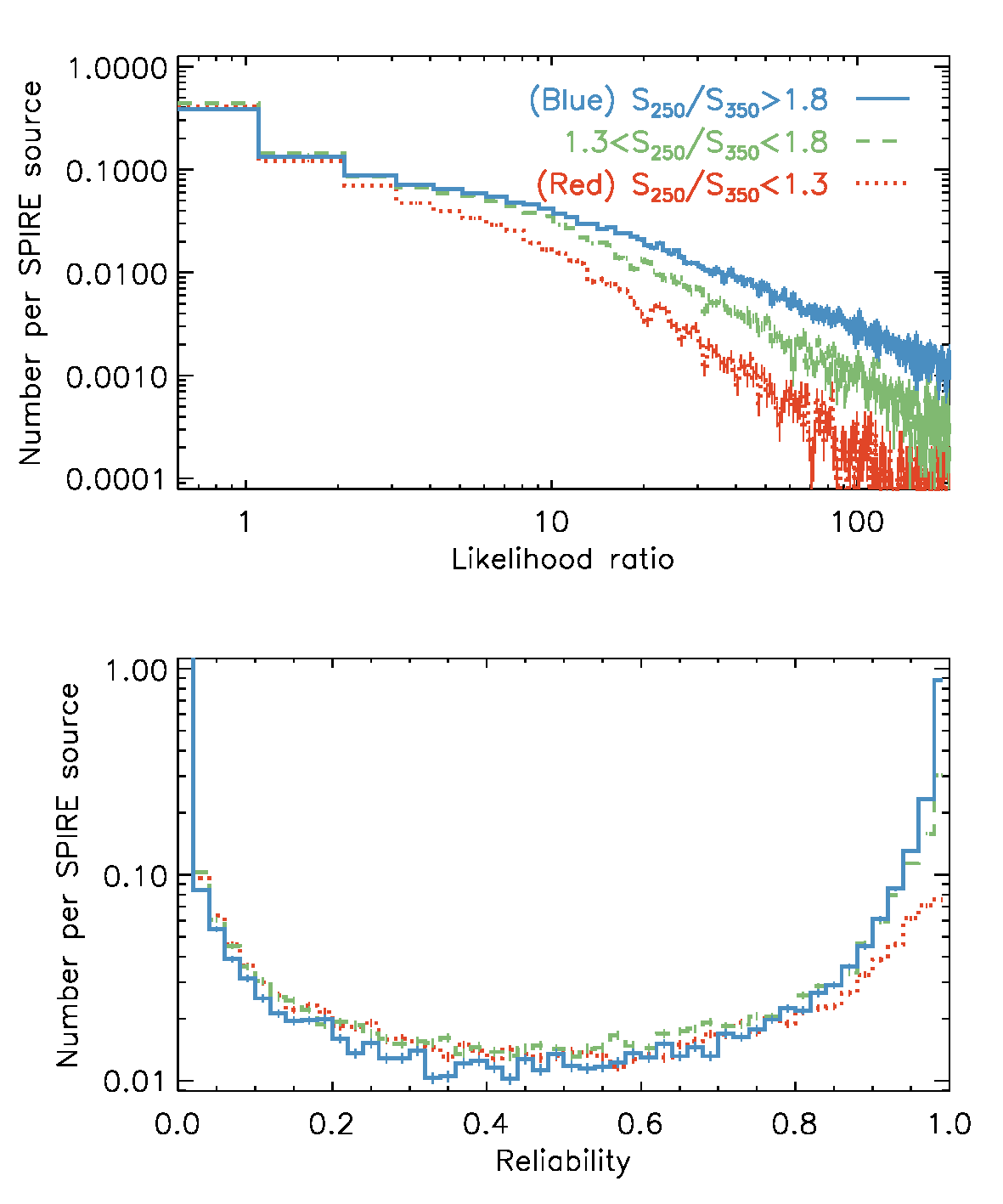}
\caption{Likelihood ratios and reliabilities of \hatlas\ sources in bins of 250/350-\mum\ colour.}
\label{fig:lr_rel_hists}
\end{center}
\end{figure}

The likelihood ratio of every potential match within 10~arcsec of each SPIRE source was calculated using \eqnref{eqn:lr}, and the reliability $R_j$ of each potential match was then computed as the ratio of its likelihood ratio $L_j$ to the sum of likelihood ratios of all potential matches in addition to the probability that there is no match; thus
\begin{equation}
R_j=\dfrac{L_j}{\sum_i L_i + (1-Q_0)}
\label{eqn:reliability}
\end{equation}
\citep{Sutherland1992}.
The reliability of a match therefore takes into account other possible matches as well as the probability ($1-Q_0$) that the true counterpart was not detected in the optical survey.
The likelihood ratios and reliabilities of all potential counterparts are shown in Fig.~\ref{fig:lr_rel_hists} for three bins of submm colour. These show that the fraction of SPIRE sources with high-reliability counterparts is much higher for blue SPIRE sources than for red ones, which may be explained by the increased probability that a red SPIRE source lies at high redshift, and/or by the increased probability that a red SPIRE source is the result of a blend between two galaxies (both of which would therefore have lower reliability than if there were only one). The figure also shows that SPIRE sources of all colours have a large number of low-reliability matches which are mostly likely to be chance alignments or correlated galaxies, but could still contribute to confusion noise in the submm (if they are submm emitters) since these matches are all within 10~arcsec.

\begin{table*}
\caption{Results of ID analysis by field}
\begin{center}
\begin{tabular}{l r r r r}
\hline
\hatlas\ sample & GAMA9 & GAMA12 & GAMA15 & Total\\
\hline
SPIRE SNR$_{250} \geq 4$ & 37612 & 36906 & 39479 & 113997\\
SDSS candidates $\leq10\arcsec$ & 42770 & 40462 & 44898 & 128130\\
Reliable IDs ($R\geq 0.8$) & 14280 & 14900 & 15655 & 44835\\
Stars 				 & 95       & 224     & 103 & 422\\
Galaxies 				 & 13864 & 14389 & 15199 & 43452\\
QSOs (spectroscopic)	 &  159    &  203    & 248 & 610\\
QSOs (photometric)		 &   162   &  84      & 105 & 351\\
Spec-$z$				 &  7422  &  7727  & 8589 & 23738\\
GAMA $FUV$ detections   & 3618 & 4089 & 4698 & 12405\\
GAMA $NUV$ detections   & 5344 & 5543 & 6449 & 17336 \\
GAMA $r$ detections	  & 7472 & 7655 & 8371 & 23498 \\
GAMA $K$ detections	  & 7494 & 7669 & 8381 & 23544 \\
GAMA $W1$ detections      & 7168 & 7457 & 8116 & 22741 \\
GAMA $W4$ detections      & 1587 & 1963 & 2786 & 6336 \\
\hline
\end{tabular}
\end{center}
Note: photometric detections are defined as SNR~$\geq3$ in the GAMA photometric catalogue \citep{Driver2015}. 
All counterparts are detected in SDSS $r$ by definition, but GAMA data are available only for counterparts with $r<19.8$ which are classified as galaxies by \citet{Baldry2010}.
\label{tab:results}
\end{table*}

\subsection{Catalogue checks and flags}
\label{sec:checks}
The LR procedure relies on the assumptions that SPIRE sources are unresolved at 250\mum, that the optical positions and magnitudes are precisely known, and that the optical catalogue is complete. These assumptions break down in a minority of cases. Large, nearby galaxies are resolved in the SPIRE image so the positional error on the SPIRE source does not follow $f(r)$, and the positional error of the optical centroid is also non-negligible. Furthermore, the optical magnitude can be unreliable due to limitations in the automated SDSS photometry especially in the case of clumpy star-forming discs or discs with dust lanes, and clumpy galaxies or bright stars can be broken up into multiple sources in SDSS.\footnote{Note that such problems are less likely to affect the GAMA photometry which use methods better optimised for galaxies.} 
The LR and reliability results can under-estimate the true reliability of a cross-match in such cases.
It is also unavoidable that optical IDs are missed due to incompleteness in the optical catalogues, especially where SDSS catalogues are masked close to bright stars.

While we cannot realistically hope to correct all such failures in the catalogue, we should at least ensure that no bright SPIRE sources are missing IDs that would be easily identifiable by eye. 
We therefore visually inspected the brightest 300 SPIRE sources in each of the three fields, and flagged objects with missing optical counterparts, with mis-classified optical counterparts, or where the reliability was under-estimated for resolved SPIRE sources or for galaxy mergers.
We also visually inspected all SPIRE sources within 10 arcsec of a star in the Tycho catalogue, to ensure that no bright stellar IDs were missing from SDSS, and found that none were missing.
A total of 13 missing IDs were added as a result of visual inspection, taking data from the NASA/IPAC Extragalactic Database (NED), GAMA and SDSS. We also modified the reliabilities of 65 IDs which had been under-estimated by the automated procedure (resolved galaxies and mergers), changed the galaxy/star classification of 40 bright objects that had been misclassified (mostly bright stars with saturated optical photometry), and removed 28 optical counterparts that were found to be artefacts or duplicates resulting from the breaking up of large galaxies or bright stars in the SDSS catalogue (in general these had not been given high reliability, but occasionally they affected the reliability of the true counterpart). Since all of these problems are associated with bright objects we are confident that their effect on the IDs of fainter sources that were not inspected is small or negligible.
In addition to those visually inspected, we searched SDSS DR8 for any matches within 10 arcsec of SPIRE sources with no potential counterpart in the original optical catalogue (leading to 1105 additional optical candidates, all of which were visually confirmed).\footnote{Full details of all additions and flags are provided as part of the data release.}
The LR and reliability values were recalculated for all potential counterparts to SPIRE sources affected by these additions.
The numbers in Table~\ref{tab:results} reflect these changes, but they are not included in the input samples described in Table~\ref{tab:stats}.

\subsection{`Reliable' IDs: completeness and contamination}
\label{sec:comp}
Details of the IDs in each field are listed in Table~\ref{tab:results}.
As in \citetalias{Smith2011a}, we choose to define `reliable' counterparts as all matches with $R \geq 0.8$, in order to select a sample which is both very clean (minimizing false counterparts) and dominated by sources with low blending.
This latter result is due to the $L_i$ values on the demoninator of \eqnref{eqn:reliability}. 
For example, if a close pair of galaxies with similar magnitudes are found at equal distances from the submm centroid, and both have very high likelihood ratios ($L_1\approx L_2 \gg 1-Q_0$), either galaxy could be identified as a likely counterpart. However, the high likelihood ratio of each one reduces the reliability of the other since the ratio $L_1/(L_1+L_2)\approx0.5$  ensures that the reliability of each is not greater than 0.5. 
Thus a cut of $R \geq 0.8$ is biased against potentially interacting galaxies, including blended submm sources as well as interacting systems where only one object is a submm source; on the other hand it leads to well-matched fluxes for spectral energy distribution (SED) fitting \citep[\citetalias{Smith2011a};][]{Smith2012c}.

The contamination rate of the `reliable' sample can be estimated from the assumption that the probability of a match being false is given by $1-R_j$, so that the total number of false IDs with $R\geq 0.8$ is
\begin{equation}
N_\text{false}=\sum\limits_{R \geq 0.8} (1-R)
\label{eqn:contamination}
\end{equation}
(\citetalias{Smith2011a}). Hence there are 2101 false `reliable' IDs in DR1, or $4.7$ per cent, in comparison with $4.2$ per cent in SDP. The difference is slightly larger than the Poisson errors on these percentages (0.1 and 0.4 respectively) and is unlikely to be due to differences in the SDSS sample because the same magnitude limits were used in both cases. The discrepancy could be due to cosmic variance, or it could be due to the change in the flux limit of the SPIRE catalogue. The average 4-$\sigma$ limit of 28\,mJy in DR1 is slightly deeper than the 32\,mJy limit in SDP, hence we could suffer greater contamination from high-redshift submm sources falsely identified with low-redshift SDSS objects. Finally, this estimate of the contamination is also sensitive to the measurements of $Q_0$ and \sigmapos, which have changed between SDP and DR1.

The completeness of the `reliable' sample is calculated as
\begin{equation}
\eta=\dfrac{n(R\geq0.8)}{n(\text{SNR}_{250}\geq4)} \dfrac{1}{Q_0},
\label{eqn:completeness}
\end{equation}
i.e. simply the number of reliable IDs divided by the estimated true number of sources within both the $r$ and 250-\mum\ limits (\citetalias{Smith2011a}), giving $\eta=73.0$ per cent. 
The DR1 reliable sample therefore appears more complete than the SDP one ($\eta=61.8$ per cent). 
The completeness of reliable $r<22.4$ IDs in DR1 is similar to the completeness of $K_s<19.2$ IDs in GAMA9 (70 per cent) found by \citet{Fleuren2012}, although the average 250-\mum\ flux limit used here is 
4\,mJy deeper, and the $Q_0$ value is lower (0.54 versus 0.73 in \citeauthor{Fleuren2012}), hence the comparison is not direct.

\begin{figure}
\begin{center}
\includegraphics[width=0.48\textwidth,clip,trim=9mm 3mm 3mm 6mm]{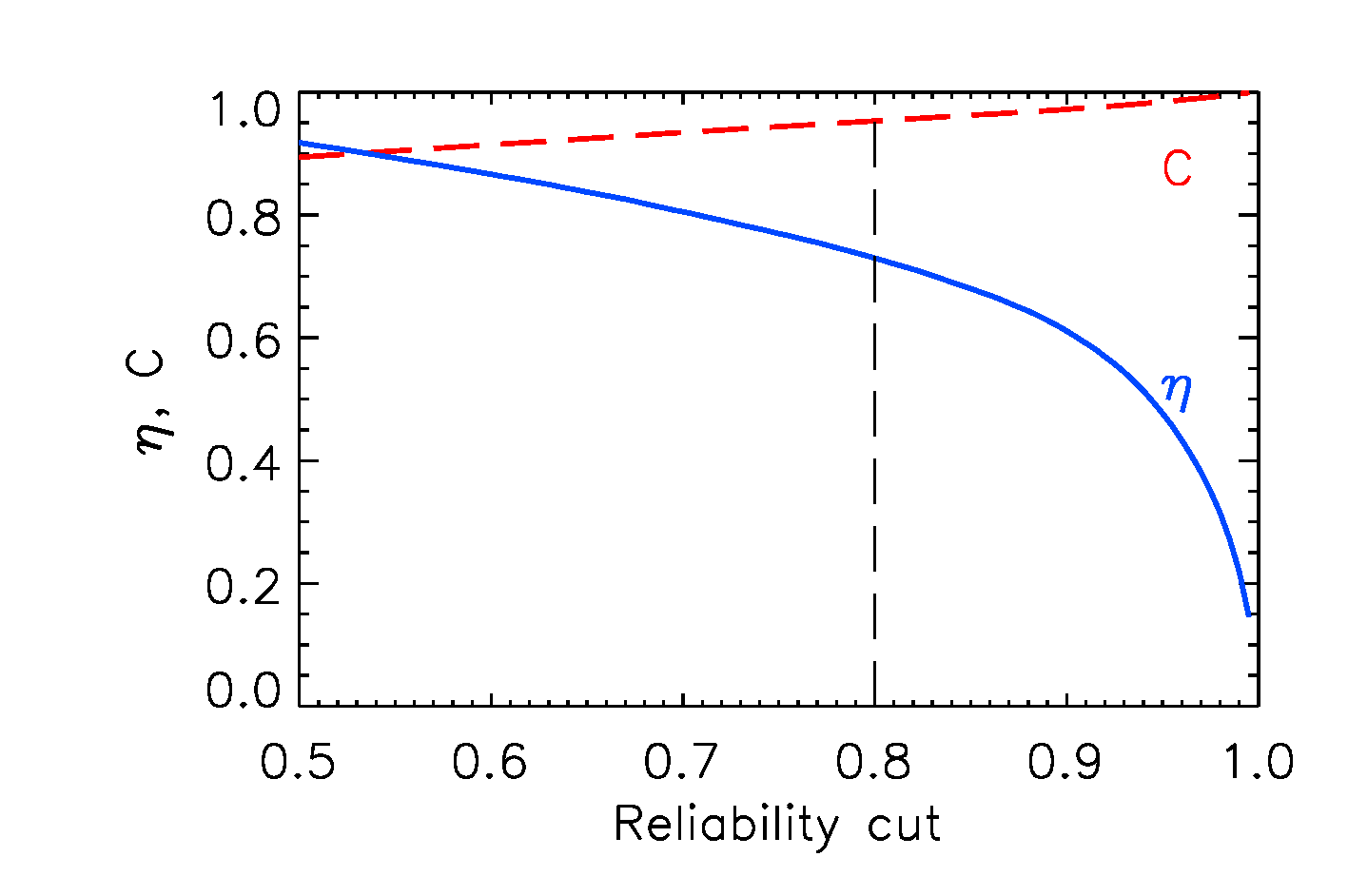}
\caption{The effect of changing the reliability cut on the `reliable' sample's completeness $\eta$ (blue solid line), given by \eqnref{eqn:completeness}, and cleanness (red dashed line), given by $C=1-N_\text{false}/N_\text{SPIRE}$. The vertical dashed line indicates the cut used to define the `reliable' sample in this and other \hatlas\ papers. In principle, any value can be chosen for this cut to suit the needs of a given project.}
\label{fig:relcuts}
\end{center}
\end{figure}

We can assess the effect of changing the reliability cut by measuring the completeness and cleanness of the `reliable' sample as a function of the threshold, as shown in Fig.~\ref{fig:relcuts}. The chosen cut of $R\geq 0.8$ selects a very clean sample at the expense of completeness. Any increase in the cut would lead to a sharp fall in completeness, although a lower cut could be used to increase completeness with a modest drop in cleanness. However, it is not possible to get close to 100 per cent completeness without a cut of around 0.5 or less, and such a cut would lead to complications in the analysis since a one-to-one matching between SPIRE and optical counterparts would not be guaranteed. Note that 100 per cent completeness would mean that the fraction of sources with reliable IDs is equal to $Q_0$, not unity.

We have chosen to use the cut of $R\geq 0.8$ for our analysis, but the data release contains the full set of potential counterparts with likelihood and reliability data for each, so users are free to make their own decision on this issue.

\subsection{Sensitivity to the value of Q$_\textbf{0}$}
\label{sec:lr-q0}
Our assumption of a universal constant $Q_0$ may lead to inaccuracies for certain subsets of sources since the fraction of real IDs which appear in SDSS will be a function of redshift, hence will vary with both submm flux and colour.
In Section~\ref{sec:offsets} we estimated $Q_0$ from the SPIRE--SDSS positional offset histograms for blue ($S_{250}/S_{350}>2.4$) SPIRE sources in SNR bins, and found higher $Q_0$ for the blue sources, increasing with SNR (Table~\ref{tab:params}). Although these values could be slightly biased by multiplicity of counterparts, we can use them to test the sensitivity of the LR results to $Q_0$. The advantages of using this blue subset are that it is  unlikely to contain a significant lensed population which could bias $Q_0$, and that it is likely to have the largest deviation from the average $Q_0$, since the majority of the sources will be at low enough redshifts to appear in SDSS.
The average $Q_0$ for blue SPIRE sources and galaxy/QSO candidates in SDSS is 0.858 (the number-weighted average of values in Table~\ref{tab:params}). 

We re-calculated the LR for the 13992 blue SPIRE sources using $Q_0=0.858$ for extragalactic objects (keeping $Q_0=0.020$ for stars unchanged, on the assumption that their $Q_0$ is independent of SPIRE colour), and compared the results to the release catalogue which was processed using $Q_0=0.519$ for extragalactic objects.
The release catalogue contains reliable IDs for 10320 of the 13992 blue SPIRE sources, but using the higher $Q_0$ leads to an additional 708 IDs, or 5.1 per cent of the blue SPIRE sources.
If we assume that the true value of $Q_0$ for this subset is $0.858+0.020$, then \eqnref{eqn:completeness} indicates that the completeness of IDs for blue sources in the release catalogue is 84 per cent, but this would rise to 90 per cent if we were to include the additional 708 IDs.

Extending this analysis to the whole population of SPIRE IDs is not straightforward, but our underestimate of $Q_0$ is certain to be greatest in this blue SPIRE bin, hence overall statistics will be affected less significantly than the results above indicate. $Q_0$ for the reddest sources will be lower than the average (they are at higher redshift and less likely to appear in SDSS), and the bias shown above is likely to be reversed for that subset. However, we cannot reliably measure $Q_0$ for red sources unless we can account for the bias from lensed sources which are falsely identified with the foreground lensing system in SDSS. 
For this reason we have used the overall estimate of $Q_0$ in our release catalogue. 
A more ideal solution would be to measure $Q_0$ as a function of SPIRE flux and colour, accounting for the bias from lensing, but we leave this extension to the method for future work.

\subsection{Sensitivity of stellar counterparts}
\label{sec:stellarids}
In Section~\ref{sec:qm} we described a change in the method used for calculating the LR of stellar counterparts, by assuming a constant $q(m)/n(m)$ instead of a constant $q(m)$ (as in \citetalias{Smith2011a}), thereby discarding the assumption of a link between optical magnitude and detection in the submm. Our motivation for this choice was based on the high measured $q(m)/n(m)$ at faint magnitudes and the weakness of the presumed link between optical and submm brightness.
The effect of this decision is significant for the statistics of stellar IDs. By assuming higher $q(m)/n(m)$ at faint $m$ we boost the LRs of the majority of stars, at the expense of the few bright stars, and this in turn leads to higher reliabilities and an overall count of 422 reliable stellar IDs. For comparison, the flat-$q(m)$ method employed in SDP (\citetalias{Smith2011a}) would lead to only 161 reliable stellar IDs in DR1. This also impacts on the galaxy IDs, because their reliabilities depend on the LRs of all potential counterparts including stars, and so our method yields 188 fewer reliable galaxy IDs than the flat-$q(m)$ method.
The completeness and contamination fractions of galaxy IDs do not change significantly, but the estimated completeness of stars is higher using our new method ($\eta=18.1$~per cent versus $7.1$~per cent), although the contamination from false IDs is also higher ($14.9$~per cent versus 10.3 per cent) -- see equations (\ref{eqn:contamination}) and (\ref{eqn:completeness}).
Furthermore, the number of stellar IDs in common between the two methods is only 41, meaning that the reliability of stellar counterparts is highly dependent on the assumptions made about $q(m)$. Since we cannot measure this at $m<21.5$, it is not necessarily true that our new method is an improvement on the SDP method, although it is consistent with the data that we have for $q(m)$, hence the reliabilities of stellar IDs must be taken with a degree of caution (as previously noted by \citetalias{Smith2011a}).

A further degree of uncertainty arises from the fact that the relatively high $q(m)$ measured for stars at $m>21.5$ (see Fig.~\ref{fig:qm} middle panel) may be biased by mis-classification of faint galaxies as stars based on SDSS imaging. If this contamination of the stars is significant then we may have over-estimated the LRs of stellar counterparts [due to over-estimating $q(m)/n(m)$], while those mis-classified galaxies will have their LR underestimated [since galaxies have higher $q(m)/n(m)$]. This issue can only be resolved with further investigation of the statistical properties of submm-bright stars.

In spite of these uncertainties, there are likely to be many more stars detected in \hatlas\ than the reliable sample would suggest (as indicated by the low completeness), although the LR process has failed to identify unambiguously which sources they are, owing to inadequate prior knowledge of the magnitude distribution of 250-\mum-detected stars. This situation could perhaps be improved by stacking analyses or prior-based `forced photometry' on known star positions, or an analysis of the \emph{Herschel} debris disc surveys DUNES \citep{Eiroa2010} and DEBRIS \citep{Matthews2010}.
We leave such improvements for future work.

\subsection{The incidence of multiple counterparts}
\label{sec:multiplicity}
The disadvantage of the reliability metric is that it assumes each SPIRE source has a single counterpart, although this counterpart may not be detected in the optical survey. This means that the reliability of a given match can be inaccurate if the SPIRE source is a blend of two or more galaxies, all of which could be treated as genuine counterparts.
The bias against interacting systems (mentioned in Section~\ref{sec:comp}) is especially relevant since many bright far-IR sources are known to be interacting \citep[e.g.][]{Bettoni2012}.

It is difficult to reliably estimate the fraction of SPIRE sources which have multiple genuine counterparts. 
One can start by comparing the results of using the likelihood ratio (LR) instead of the reliability to define counterparts (\citetalias{Smith2011a}), since the LR contains no information on the existence of other potential counterparts but simply gives the ratio of the probability of a single potential match to that of a chance alignment. In cases where there is only one potential ID, \eqnref{eqn:reliability} implies that an LR of $1.924$ corresponds to the threshold $R=0.8$ for galaxy matches ($Q_0=0.519$). Hence, to avoid the effects of multiplicity on the reliable IDs we could apply this LR cut and find 50421 extragalactic IDs, of which 6539 (13.0 per cent) fail the $R\geq0.8$ threshold. These additional IDs could be considered candidates for sources with multiple counterparts which have been missed by the reliability cut, although in reality these will be a mixture of true multiples (mergers, pairs and close groups) as well as chance alignments where the single true counterpart is ambiguous.

An alternative approach is to suggest that multiple counterparts are missed by the requirement for a single match to have $R\ge0.8$, but can be recovered by assuming that the sum of their reliabilities meets this threshold \citep{Fleuren2012}. In DR1, 3483 SPIRE sources have multiple extragalactic counterparts whose $\sum R\geq0.8$, but for which no single counterpart meets the threshold. This estimate for the missed multiple counterparts is smaller than that from the LR threshold (6539), because there will be systems where multiple low-reliability (but high-LR) counterparts do not have a high combined reliability.
Following \citet{Fleuren2012}, we can further clean this list of candidate multiple counterparts by selecting only those with spectroscopic redshifts which agree within five per cent, or photometric redshifts within ten per cent, in order to exclude chance alignments. This requirement reduces the number to 2071 SPIRE sources with multiple extragalactic counterparts at the same redshift. This is a small fraction of the total number of reliable IDs, so we can conclude that most SPIRE sources in \hatlas\ have single galaxy IDs.\footnote{Note that the fraction of submm sources with multiple IDs will depend on the flux density regime being considered and on the redshifts of the sources.}

A related issue is the incompleteness of the sample due to multiplicity, which results from the fact that additional candidates in the search radius reduce the reliability of the true counterpart.
\citet{Fleuren2012} estimated this incompleteness by comparing the fraction of reliable IDs among all SPIRE sources with one candidate, those with two candidates, and so on. In Table~\ref{tab:multiplicity} we show these fractions for SPIRE 250-\mum\ sources with between 0 and 10 candidates, including either all candidates or only extragalactic ones. For example, 49057 SPIRE sources have only one potential counterpart within 10~arcsec from the full optical catalogue, and of these 52 per cent are reliable IDs; 24194 have two potential counterparts but of these 58 per cent are reliable, suggesting that sources are more likely to be assigned an ID if there are two candidates in the search radius. The fraction of reliable counterparts then falls for increasing numbers of potential candidates. Looking at extragalactic candidates only, the fraction of reliable counterparts is 61 per cent when there is only one candidate, and this fraction falls when more candidates are available. One might ask why the fall was not seen when counting all candidates, and we suggest that the reason is that the reliable fraction of sources with only one candidate ID is depressed by the fact that a disproportionate number of stellar IDs fall in this bin (stars tend to be isolated), and their reliabilities are generally lower.

The falling reliable fraction with increased number of extragalactic candidates can be interpreted as incompleteness resulting from multiplicity. We can estimate this by adding up the total number of missing `reliable' counterparts from each row of the table, assuming that in each case the intrinsic number is 61 per cent of $N_\text{SPIRE}$, and the result is 408 missing IDs. Hence we conclude that the incompleteness of IDs resulting from multiplicity is small.

\begin{table}
\caption{Number of SPIRE sources and the subset of those with reliable IDs, as a function of the number ($p$) of candidate IDs within the 10-arcsec search radius [counting either all candidates or extragalactic (xgal) candidates only].}
\begin{center}
\begin{tabular}{r r r r r}
\hline
   &   \multicolumn{2}{c}{With $p$ candidates:} & \multicolumn{2}{c}{With $p$ xgal candidates:}\\
   $p$  &  $N_\text{SPIRE}$  & \multicolumn{1}{c}{$N_\text{rel}$}  & $N_\text{SPIRE}$  & \multicolumn{1}{c}{$N_\text{rel}$} \\
\hline
       0 &      31394 &     0~~~(0\%)          &  41142 &    0~~~(0\%)      \\
       1 &      49057 &     25648      (52\%) & 49898 &    30669      (61\%)  \\
       2 &      24194 &     14026      (58\%) & 18013 &    11043      (61\%)   \\
       3 &       7365  &      4075      (55\%)  & 4109 &  2245      (55\%) \\
       4 &       1661  &       897      (54\%)   & 722 &   366      (51\%)  \\
       5 &        280   &      123      (44\%)    & 96 &    41      (43\%)   \\
       6 &         39    &      18      (46\%)      & 12 &    5      (42\%)  \\
       7 &          4     &      2      (50\%)        & 2 &    1      (50\%) \\
       8 &          3     &      1      (33\%)        & 3 &      1      (33\%) \\
       9 &          0     &      0~~~~~(--)         & 0 &         0~~~~~(--)  \\
\hline
\end{tabular}
\end{center}
\label{tab:multiplicity}
\end{table}

\section{Optical properties of \hatlas\ IDs in SDSS}
\label{sec:optresults}
In this section we explore the properties of \hatlas\ sources with `reliable' optical counterparts,
using optical photometry from SDSS and redshifts from the sources described in Section~\ref{sec:z}. 

\subsection{Redshifts and magnitudes}
\label{sec:zdist}
\begin{figure}
\begin{center}
\includegraphics[width=0.48\textwidth,clip,trim=12mm 3mm 0mm 3mm]{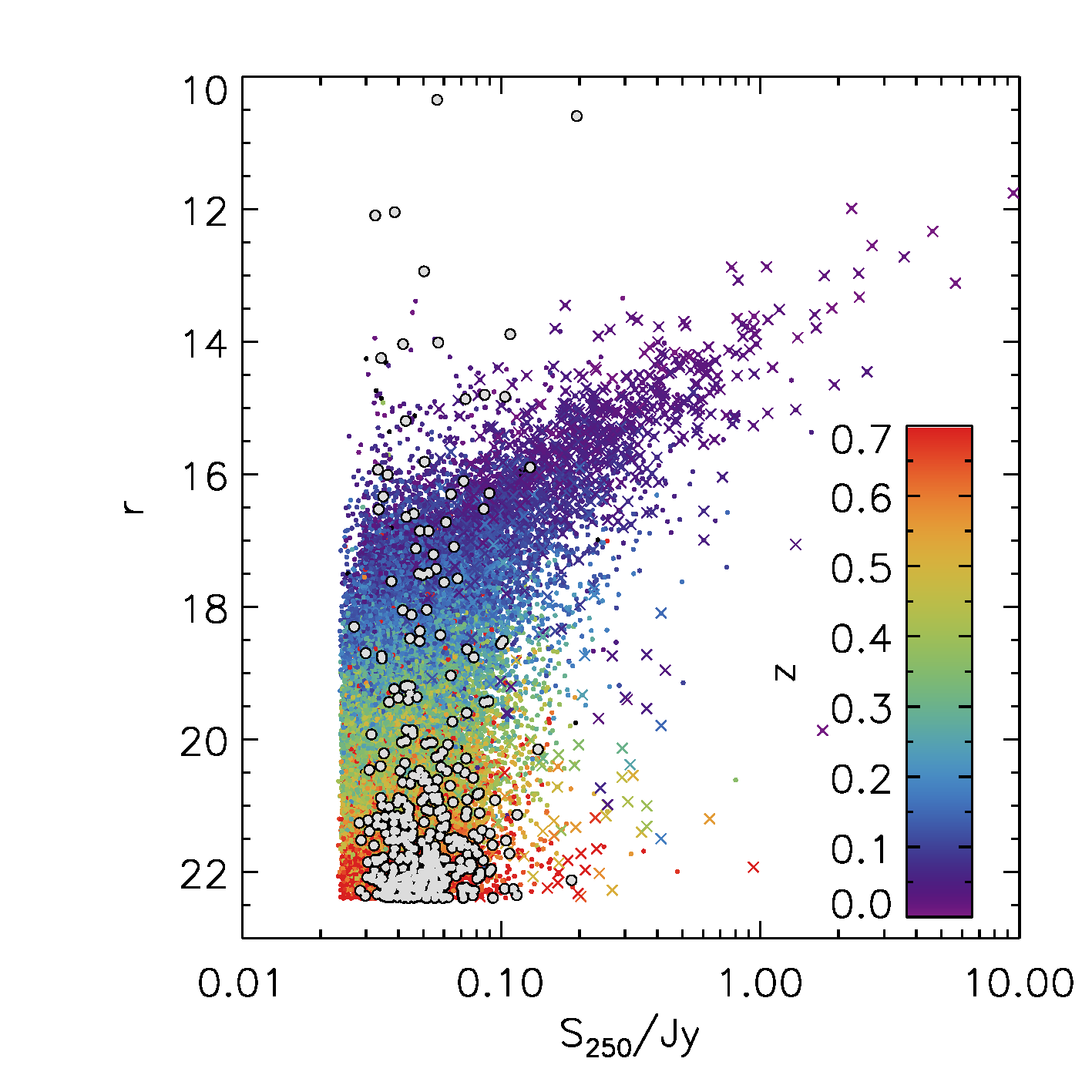}
\caption{Optical magnitudes in the $r$ band versus 250-\mum\ fluxes of \hatlas\ sources with reliable IDs in SDSS. Extragalactic sources are coloured according to their redshift (spectroscopic where available; photometric otherwise). Stars are shown by grey circles. 
For most sources we plot the point-source flux at 250\mum, which is the flux measurement used to define the 4-$\sigma$ sample. However, for a relatively small number of extended sources at low redshifts (plotted as crosses) we plot the aperture flux as described in \citetalias{Valiante2014}, where this exceeds the measured point-source flux.}
\label{fig:r-250}
\end{center}
\end{figure}

Figure~\ref{fig:r-250} shows the relationship between optical magnitude of all reliable SDSS counterparts, 250-\mum\ flux and redshift (for extragalactic counterparts). The 250-\mum\ flux is weakly correlated with redshift, in contrast with the optical magnitude which is a strong function of redshift; this is a result of the differential $k$-corrections. Almost all of the bright sources in \hatlas\ with $S_{250}\gtrsim 200$\,mJy are at $z<0.1$, but at lower fluxes, sources are detected at all redshifts up to $z\sim0.7$ and the upper limit in the redshift distribution is clearly a result of the depth of SDSS rather than that of \hatlas.

Stellar counterparts, shown by grey circles on Fig.~\ref{fig:r-250}, 
exhibit no correlation between their optical and submm fluxes.
As discussed in Section~\ref{sec:qm}, we do not necessarily expect a correlation between the optical and submm fluxes of the stars we can detect, although we note that the increased number of stellar IDs towards fainter optical magnitudes $r\gtrsim 21$ may be a cause for concern. It is likely that the optical classification of stars is more uncertain close to the $r$-band limit, and that more galaxies are mis-classified as stars. If the large number of stellar IDs at $r\gtrsim21$ in  Fig.~\ref{fig:r-250} are in fact mis-classified galaxies, this may indicate a larger number of such mis-classifications in the optical catalogue, and since the $Q_0$ for stars is much lower than that for galaxies, these objects are much less likely to have been assigned reliable IDs. Hence the completeness of galaxy IDs at faint $r$-band magnitudes may be adversely affected.
In fact we can estimate how bad this problem could be by looking at the completeness as a function of magnitude estimated firstly from all possible counterparts, and secondly from only those potential counterparts that we have classified as extragalactic (we describe how this is measured in Section~\ref{sec:compz}). The completeness at $r>21$ falls by roughly 0.05 between these two samples (at brighter magnitudes it doesn't change significantly).  This means that even if all the star classifications at $r>21$ are false, and should be galaxies, then our completeness is only about 5 per cent worse than it should be at these faint magnitudes. On the other hand, the contamination of stellar IDs by unresolved galaxies could be significant (as mentioned in Section~\ref{sec:stellarids}), and optical imaging at higher resolution (compared with typical SDSS seeing of 1.5~arcsec) is needed to confirm these.

Approximately half (49 per cent) of reliable IDs have a spectroscopic redshift, while the overall fraction of all potential counterparts in the catalogue which have spectroscopic redshifts is 21 per cent. 
These redshifts originate primarily from GAMA (69 per cent) and SDSS DR7/DR10 (24 per cent); the full breakdown of all redshifts (not only reliable IDs) is shown in Table~\ref{tab:zsource}.
2583 of the 22808 spectroscopic redshifts from GAMA were obtained within the \hatlas\ filler program (as described in Section~\ref{sec:z}).

\begin{table}
\caption{Number of good quality ($nQ\geq 3$) redshifts in the full ID catalogue originating from each survey.}
\begin{center}
\begin{tabular}{l l r}
\hline
Survey & Reference & Number \\
\hline
GAMA-AAO &\citet{Liske2015}           &      22804 \\
GAMA-LT &\citet{Liske2015}               &        4       \\
SDSS DR7 &\citet{Abazajian2009}     &       5811    \\
SDSS DR10 &\citet{Ahn2013}          	&        2424   \\
WiggleZ &\citet{Drinkwater2010}        &    1226       \\
2dFGRS &\citet{Colless2001}             &       777       \\
MGC &\citet{Driver2005}                     &     211        \\
2QZ &\citet{Croom2004}                     &    142          \\
2SLAQ-QSO &\citet{Croom2009}       &            117  \\
2SLAQ-LRG &\citet{Cannon2006}      &             59   \\
6dFGS &\citet{Jones2009}                  &       21        \\
NED & \url{http://ned.ipac.caltech.edu/} &      3           \\
UZC &\citet{Falco1999}                      &      2            \\
\hline
Total		&		&	33601 \\
\hline
\end{tabular}
\end{center}
\label{tab:zsource}
\end{table}

\begin{figure}
\begin{center}
\includegraphics[width=0.48\textwidth,clip,trim=6mm 3mm 2mm 3mm]{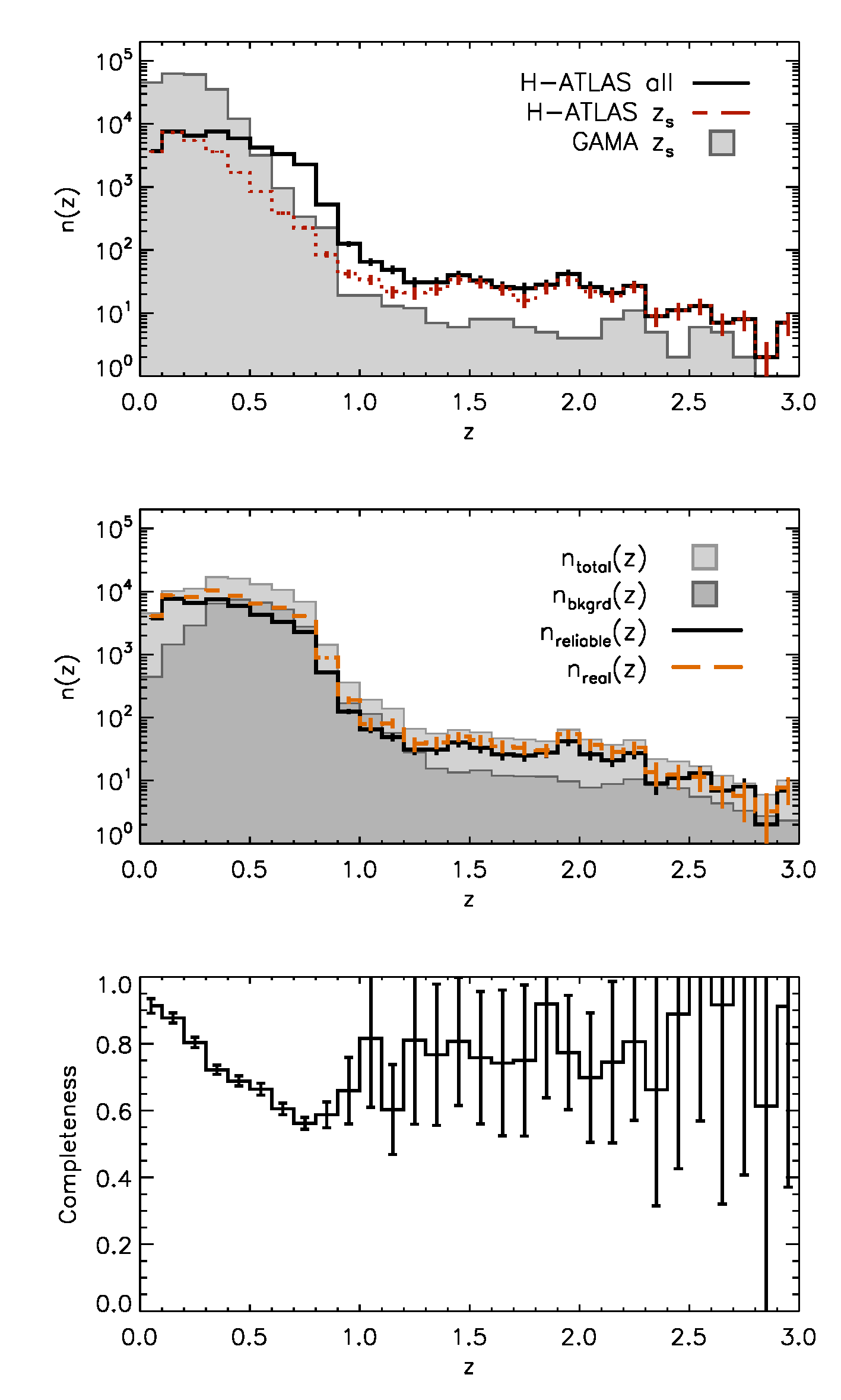}
\caption{Top: the distribution of best redshifts for the \hatlas\ reliable IDs (black solid line), and of available spectroscopic redshifts for the same sample (red dotted histogram), compared with the distribution of spectroscopic redshifts in GAMA (grey filled histogram).
Middle: the method of estimating completeness as a function of redshift from the redshift distribution of all potential galaxy matches to SPIRE sources [$n_\text{total}(z)$, light grey filled histogram], the background redshift distribution of unassociated galaxies [$n_\text{bkgrd}(z)$, dark grey filled histogram], and the redshift distribution of reliable IDs [$n_\text{reliable}(z)$, black solid line; same as in upper panel]. 
The orange dashed line shows the estimated redshift distribution of all true counterparts within the optical survey, $n_\text{real}(z)=Q_0[n_\text{total}(z)-n_\text{bkgrd}(z)]$.
Bottom: the estimated completeness given by $n_{\text{reliable}}(z)/n_\text{real}(z)$. See description in Section~\ref{sec:compz} of the text.
}
\label{fig:zdist}
\end{center}
\end{figure}

The overall redshift distribution of \hatlas\ sources is shown in the upper panel of Fig.~\ref{fig:zdist}, in comparison with that of GAMA. Our use of spectroscopic redshifts from SDSS DR10 and WiggleZ makes up for the lack of redshifts for faint galaxies in GAMA, and photometric redshifts fill in any remaining gaps. The fact that almost all reliable counterparts at $z>1$ have spectroscopic redshifts may be explained by those objects being specifically targeted by BOSS \citep{Eisenstein2011}, but it could also be affected by a bias in photometric redshifts. The potential bias arises from a tendency of the {\sc annz} code to return solutions at $z_p<1$ because this range is best represented in the training set. 
Note that Fig.~\ref{fig:zdist} only describes the redshift distribution of 250-\mum\ sources within the SDSS survey limit ($S_{250}\gtrsim 27.8$~mJy and $r<22.4$).
The unbiased redshift distribution of all \hatlas-delected sources has been investigated in more detail by \citet{Pearson2012}.

\begin{figure}
\begin{center}
\includegraphics[width=0.48\textwidth,clip,trim=6mm 3mm 3mm 3mm]{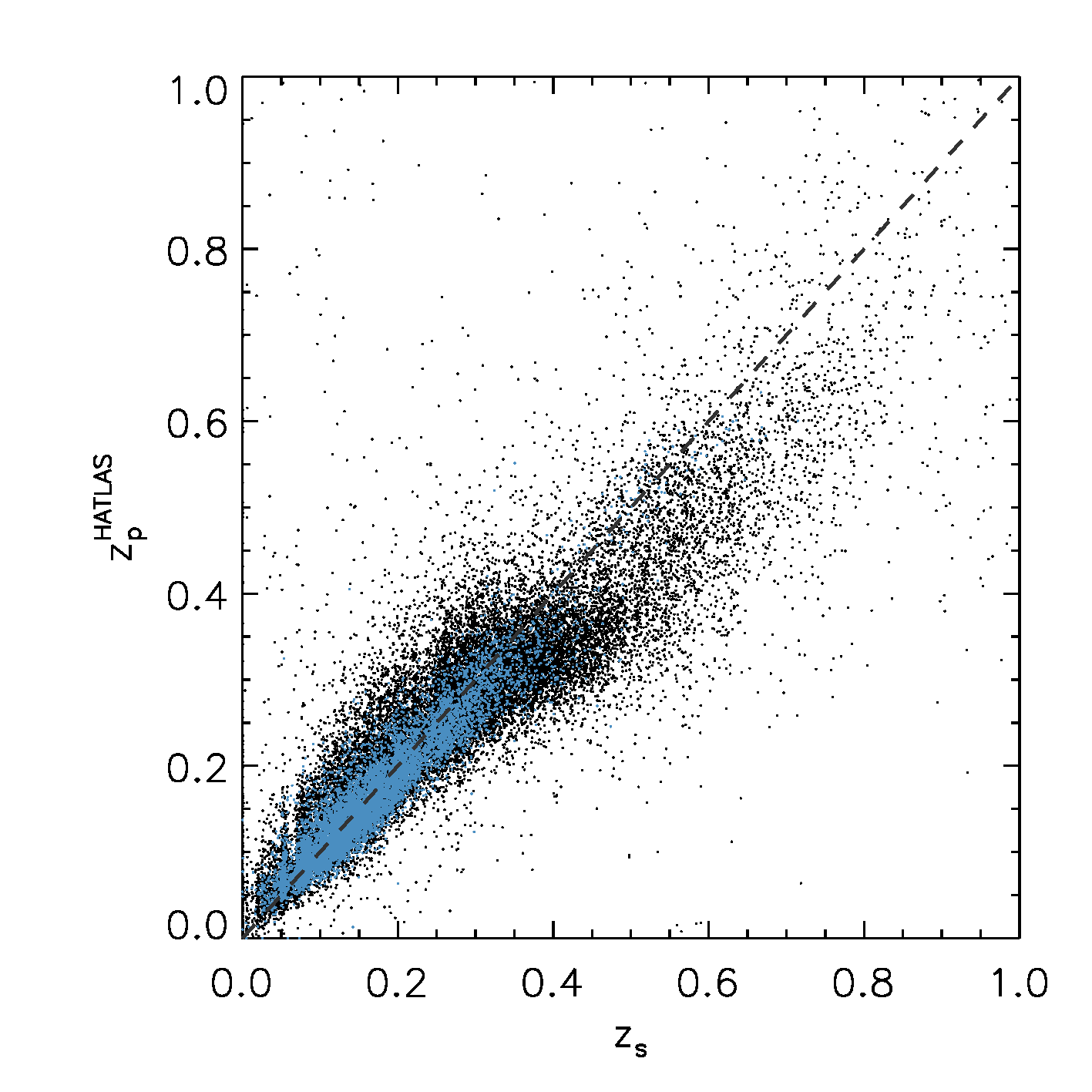}
\includegraphics[width=0.48\textwidth,clip,trim=6mm 3mm 3mm 6mm]{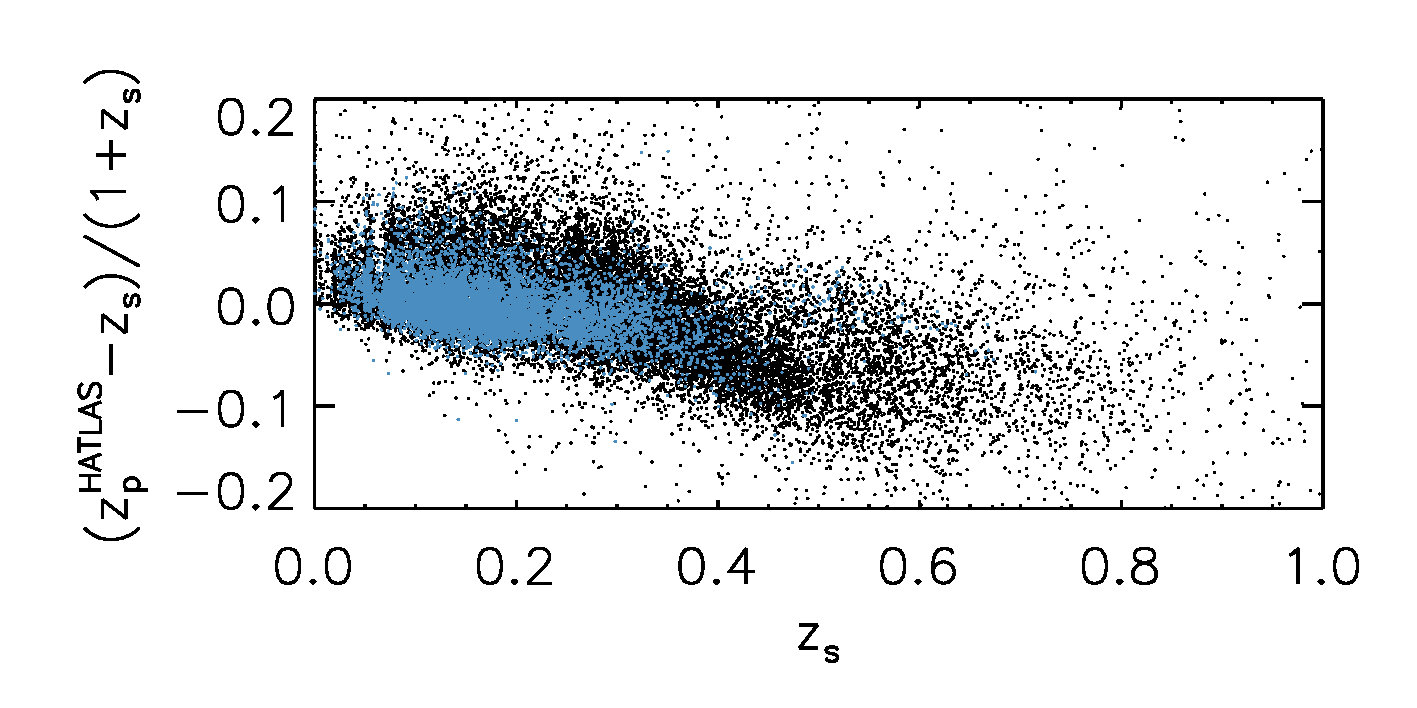}
\caption{Photometric vs. spectroscopic redshifts, as in Fig.~\ref{fig:z-z}, for reliable IDs to \hatlas\ sources. 
Black points show the $z_p$ of all IDs with a $z_s$, and blue show the subset with $z_p$ errors $\Delta z/(1+z_p)<0.02$. The lower panel shows the fractional deviations in $z_p$.}
\label{fig:z-z-ids}
\end{center}
\end{figure}

Figure~\ref{fig:z-z-ids} compares the \hatlas\ photometric redshifts to spectroscopic redshifts, for those reliable IDs which have a good spectroscopic redshift (compare this to the full set of spectroscopic redshifts in Fig.~\ref{fig:z-z}). A slight bias is seen at $z\gtrsim 0.4$, where the photometric redshifts are typically under-estimated. This appears more prominent in this reliable subset than it was for the full redshift sample (Fig.~\ref{fig:z-z}), indicating a bias in the photometric redshift algorithm to specifically under-estimate the redshifts of dusty galaxies at $z>0.4$. This tendency also existed in the SDP (\citetalias{Smith2011a} used the same photometric redshifts), and indeed was previously noted by \citet{Dunne2011}.

\begin{figure}
\begin{center}
\includegraphics[width=0.48\textwidth,clip,trim=9mm 3mm 3mm 3mm]{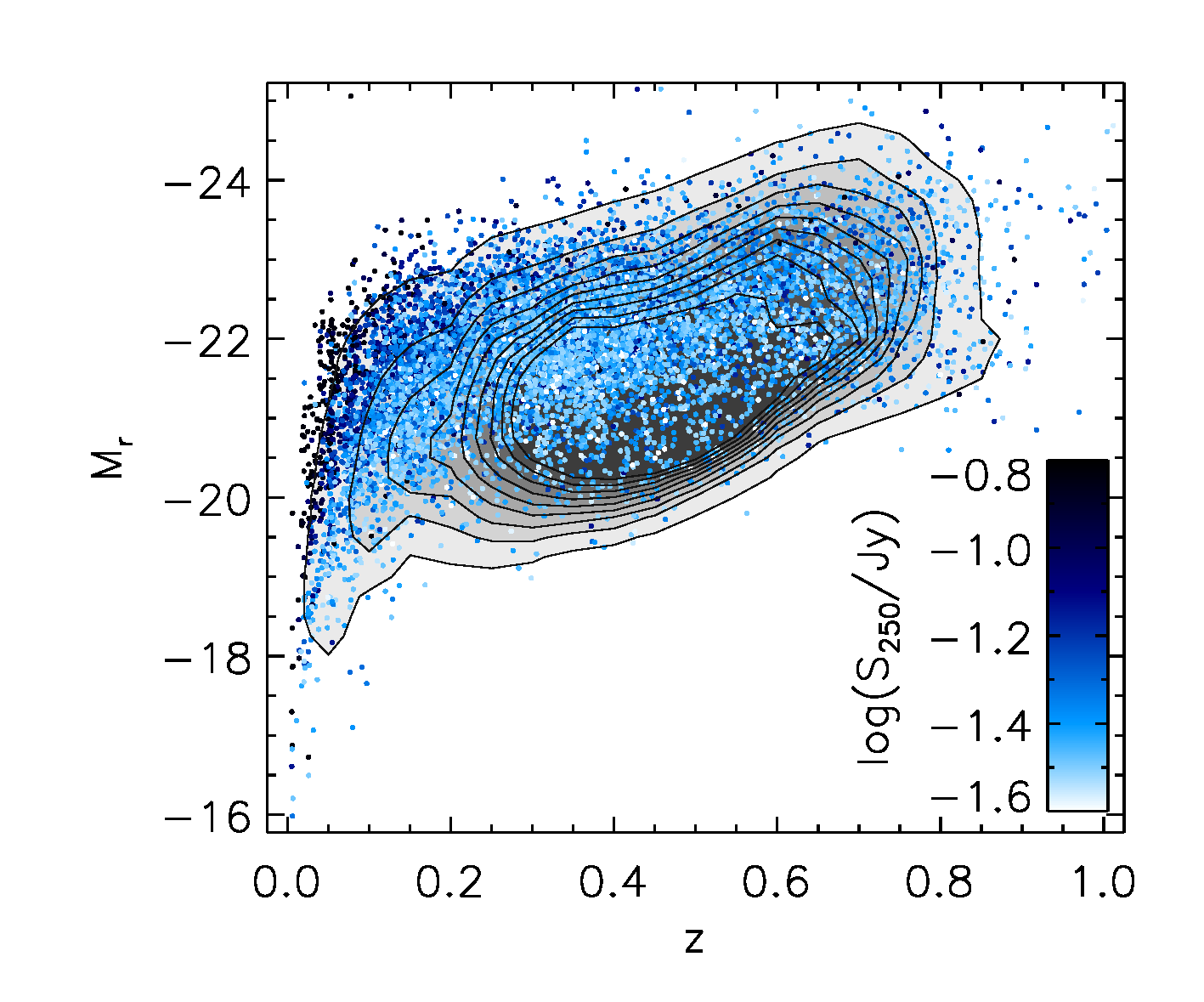}
\caption{Optical absolute magnitudes in the $r$ band as a function of redshift and 250-\mum\ flux of reliable \hatlas\ IDs (all extragalactic IDs in SDSS), with the background distribution of SDSS shown in contours. The redshifts used are spectroscopic where available, or photometric otherwise.
To reduce crowding we only plot one quarter of the reliable IDs.}
\label{fig:zmr}
\end{center}
\end{figure}

In Fig.~\ref{fig:zmr} we show the absolute magnitude $M_r$ as a function of redshift for the reliable sample. Absolute magnitudes were estimated from the SDSS photometry for the full sample using {\sc kcorrect} \citep{Blanton2007}. The \hatlas\ IDs sample the upper half of the absolute magnitude distribution in SDSS.
Flux limits above approximately 0.1\,Jy effectively sample only low redshifts ($z\lesssim 0.1$), as we saw in Fig.~\ref{fig:r-250}; the small number of higher-redshift IDs above this flux limit are likely to be strong lenses \citep{Negrello2010,Lapi2011}.
Lower fluxes span the full range of redshifts, also shown in Fig.~\ref{fig:r-250}, and submm flux is not a strong indicator of either redshift or optical luminosity.

\subsection{Completeness as a function of redshift and magnitude}
\label{sec:compz}
We can estimate the completeness of the reliable IDs sample as a function of redshift using a similar statistical analysis to that used to estimate the magnitude distributions in Section~\ref{sec:q} and in \citetalias{Smith2011a}.
The method is demonstrated on the middle panel of Fig.~\ref{fig:zdist}, in which the light grey filled histogram is the redshift distribution of all galaxies from the optical sample that are within 10~arcsec of any SPIRE position: we call this distribution $n_\text{total}(z)$. The darker filled histogram is the distribution of background galaxies that contribute to $n_\text{total}(z)$ but are not directly associated with the SPIRE sources. This distribution, $n_\text{bkgrd}(z)$, is estimated from the distribution of all redshifts in the full optical catalogue of the three fields, normalized by the ratio of the combined search area around SPIRE positions ($\pi\times10^{\prime\prime\, 2}\times N_\text{SPIRE}$) to the total field area covered by the optical catalogue. The difference $n_\text{total}(z)-n_\text{bkgrd}(z)$ therefore gives the estimated true redshift distribution of SPIRE counterparts, $n_\text{real}(z)$, which is plotted as a dashed orange line, after normalizing by $Q_0$ (the fraction of true counterparts which are within the optical catalogue).
The completeness as a function of redshift of the reliable IDs is estimated by the ratio $n_\text{reliable}(z)/n_\text{real}(z)$, where $n_\text{reliable}(z)$ is the solid black line shown in both the upper and middle panels. The completeness fraction is shown in the lower panel, with error bars based on the assumption of Poisson counting statistics.
The completeness of each redshift bin up to $z=3$ is shown in Table~\ref{tab:zcomp}.
Compared with the same statistics for SDP (\citetalias{Smith2011a}), the DR1 completeness is broadly similar, although slightly higher at $z>0.3$. This difference may be partly due to cosmic variance, and perhaps due to the greater availability of spectroscopic redshifts in the current work.
It is noticeable that the completeness remains relatively high at high redshifts, and in fact appears to be higher at $z>1$ than at $z\approx 0.7$. This may relate to a dearth of spectroscopic redshifts at $0.5<z<1$ combined with inaccuracies in photometric redshifts in this range (especially the bias discussed in the previous section). It is also possible that the probability for reliably matching a SPIRE source with its SDSS counterpart (given that it is detected in both surveys) is \emph{higher} at $z>1$ compared to lower redshifts because the magnitude-limited SDSS catalogue is less likely to contain correlated neighbours within the 10~arcsec search radius. Such neighbours would of course reduce the reliability given by \eqnref{eqn:reliability}.  Equally, the completeness at $z<0.5$ is high because the search radius corresponds to a smaller physical scale which is likely to contain fewer neighbours. 

\begin{figure}
\begin{center}
\includegraphics[width=0.48\textwidth,clip,trim=6mm 3mm 2mm 3mm]{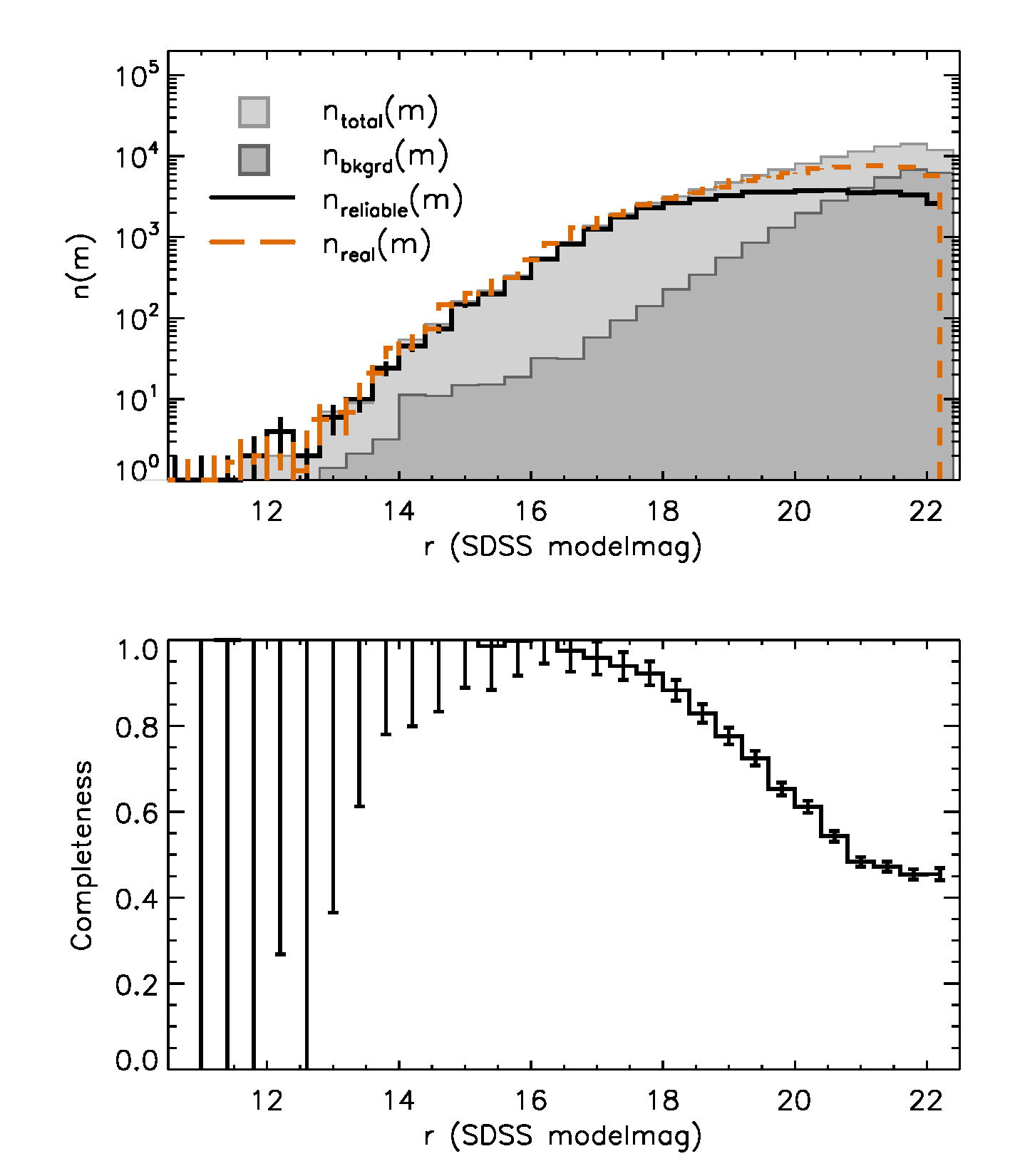}
\caption{The method of estimating completeness as a function of $r$-band magnitude from the distribution of all potential \emph{extragalactic} matches to SPIRE sources [$n_\text{total}(m)$, light grey filled histogram], the background distribution of unassociated galaxies [$n_\text{bkgrd}(m)$, dark grey filled histogram], and the distribution of reliable IDs [$n_\text{reliable}(m)$, black solid line]. The difference $n_\text{total}(m)-n_\text{bkgrd}(m)$ gives the estimated magnitude distribution of true counterparts [$n_\text{real}(m)$, orange dashed line].  
The estimated completeness given by $n_{\text{reliable}}(m)/n_\text{real}(m)$ is shown in the lower panel.
}
\label{fig:magcomp}
\end{center}
\end{figure}

A similar analysis of the magnitude histograms of true counterparts and of reliable IDs allows us to plot the ID completeness as a function of $r$-band magnitude in Fig.~\ref{fig:magcomp}. We see that the completeness is at least 90 per cent for $r<18$, but falls to around 65 per cent at $r$=19.8 (the limiting magnitude of GAMA), and is below 50 per cent for $r\gtrsim21$ (the results are tabulated in Table~\ref{tab:mcomp}).

\begin{table}
\caption{Completeness of the reliable sample as a percentage of the total number of true counterparts with $r<22.4$, as a function of redshift}
\begin{center}
\begin{tabular}{c c}
\hline
$z$ & Completeness (per cent) \\
\hline
0.0 -- 0.1 & $91.3 \pm 2.2$ \\
0.1 -- 0.2 & $87.7 \pm 1.5$ \\
0.2 -- 0.3 & $80.4 \pm 1.5$ \\
0.3 -- 0.4 & $72.2 \pm 1.3$ \\
0.4 -- 0.5 & $68.9 \pm 1.5$ \\
0.5 -- 0.6 & $66.4 \pm 1.8$ \\
0.6 -- 0.7 & $60.5 \pm 1.7$ \\
0.7 -- 0.8 & $56.2 \pm 1.8$ \\
0.8 -- 0.9 & $58.8 \pm 3.9$ \\
0.9 -- 1.0 & $65.9 \pm 9.9$ \\
1.0 -- 1.5 & $76.1 \pm 9.1$ \\
1.5 -- 2.0 & $78.9 \pm 9.9$ \\
2.0 -- 2.5 & $76.0 \pm14.0$ \\
2.5 -- 3.0 & $99.6 \pm31.4$ \\
\hline
\end{tabular}
\end{center}
\label{tab:zcomp}
\end{table}

\begin{table}
\caption{Completeness of the reliable sample as a function of $r$ magnitude, for all SDSS IDs and for extragalactic (xgal) IDs.}
\begin{center}
\begin{tabular}{c c c}
\hline
      $r$              &  All optical IDs (per cent) & xgal IDs  (per cent) \\
\hline
      10.0 -- 14.8  &    $65.2  \pm 15.6$      &      $106.3 \pm  13.0$ \\
      14.8 -- 15.2  &    $91.2  \pm 15.5$      &      $101.2 \pm  12.4$ \\
      15.2 -- 15.6  &    $96.6  \pm 14.3$      &      $98.6  \pm  10.2$ \\
      15.6 -- 16.0  &    $103.1 \pm 11.9$     &       $100.0 \pm  8.2$  \\
      16.0 -- 16.4  &    $100.8 \pm 8.0$      &       $101.0 \pm  6.4$  \\
      16.4 -- 16.8  &    $99.3  \pm 6.1$       &      $97.5  \pm  4.9$  \\
      16.8 -- 17.2  &    $94.7  \pm 4.4$       &      $95.8  \pm  3.9$   \\
      17.2 -- 17.6  &    $93.8  \pm 3.6$       &      $94.0  \pm  3.2$  \\
      17.6 -- 18.0  &    $92.4  \pm 3.0$      &       $92.2  \pm  2.7$  \\
      18.0 -- 18.4  &    $88.0  \pm 2.7$      &       $88.3  \pm  2.4$  \\
      18.4 -- 18.8  &    $83.7  \pm 2.4$      &       $82.9  \pm  2.2$  \\
      18.8 -- 19.2  &    $77.1  \pm 2.1$      &       $77.6  \pm  1.9$  \\
      19.2 -- 19.6  &    $73.2  \pm 1.8$      &       $72.5  \pm  1.7$  \\
      19.6 -- 20.0  &    $65.4  \pm 1.6$      &       $65.3  \pm  1.5$  \\
      20.0 -- 20.4  &    $60.4  \pm 1.5$      &       $61.2  \pm  1.4$  \\
      20.4 -- 20.8  &    $54.4  \pm 1.3$      &       $54.3  \pm  1.2$  \\
      20.8 -- 21.2  &    $48.6  \pm 1.2$      &       $48.3  \pm  1.2$  \\
      21.2 -- 21.6  &    $45.7  \pm 1.2$      &       $47.2  \pm  1.1$  \\
      21.6 -- 22.0  &    $42.8  \pm 1.2$      &       $45.4  \pm  1.2$  \\
      22.0 -- 22.4  &    $41.5  \pm 1.3$      &       $45.5  \pm  1.4$  \\ 
\hline      
\end{tabular}
\end{center}
\label{tab:mcomp}
\end{table}

\section{Multi-wavelength properties of \hatlas\ galaxies in GAMA}
\label{sec:multiwlresults}
In this section we explore the multi-wavelength properties of the subset of \hatlas\ sources with `reliable' counterparts in GAMA, i.e. those with $r<19.8$ which are classified as galaxies by \citet{Baldry2010}.  

\subsection{Magnitude distributions}
\begin{figure*}
\begin{center}
\includegraphics[width=\textwidth,clip=true,trim=12mm 3mm 10mm 3mm]{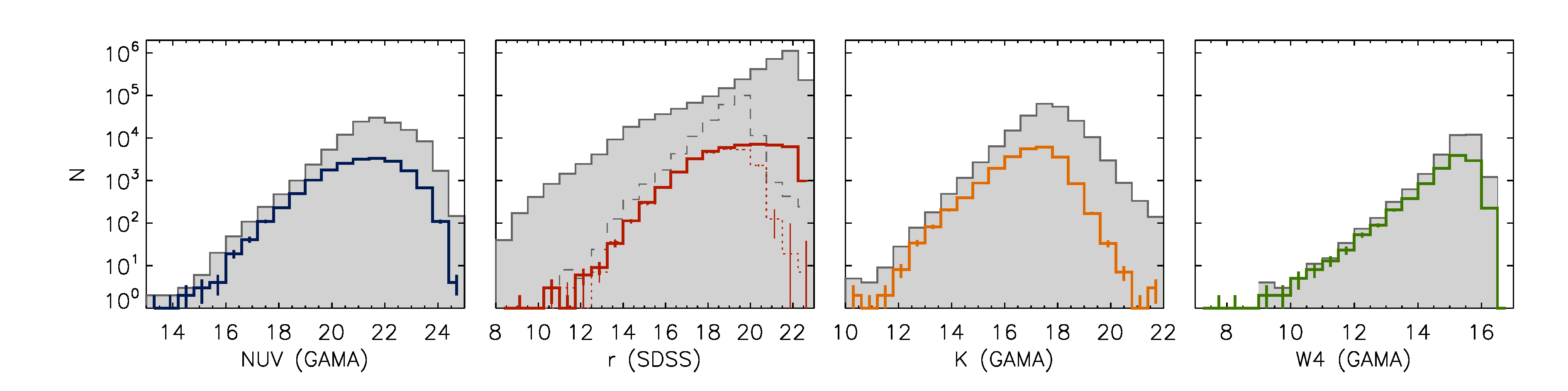}
\caption{Histograms (solid coloured lines) of the magnitudes of reliable \hatlas\ IDs in $NUV$, $r$, $K$ and $W4$ (22\mum). 
The background distributions in each band are shown by the filled histogram. The $r$-band histogram contains all SDSS objects in the optical catalogue (including stars, galaxies and quasars). The dashed grey line shows the distribution of $r$-band magnitudes in the GAMA galaxy sample. For comparison, we show the magnitude distribution of all reliable IDs in SDSS (solid red line), and that of IDs in GAMA (dotted red line).
Photometry in the other bands is drawn from the GAMA photometric catalogue only.
}
\label{fig:mdists}
\end{center}
\end{figure*}

Figure~\ref{fig:mdists} shows the magnitude distributions in four bands from the UV, optical, near-IR and mid-IR, for reliable IDs with available photometry in each band. 
For the $r$-band we can compare both to SDSS and GAMA, but for the other bands we compare to the GAMA photometric catalogue only \citep{Driver2015}. 
It is noticeable that the \hatlas\ sample selects GAMA sources in the $NUV$ down to a magnitude of around 20, and in $W4$ down to about 15, with roughly constant efficiency as a function of magnitude. This indicates that \hatlas\ is not biased against blue (UV-bright) star-forming galaxies or hot (22-\mum-bright) LIRGs (which may have been suspected due to the tendency for submm selection to favour cold dusty galaxies).
In the $r$ band, \hatlas\ detects and IDs a small fraction of all SDSS sources at all magnitudes, but a large fraction of the galaxies in GAMA at $r<18$. Most of the brighter SDSS objects missing from \hatlas\ are stars, which are relatively few in the ID catalogue. 

\subsection{Optical and infrared colours}
\begin{figure}
\begin{center}
\includegraphics[width=0.48\textwidth,clip,trim=12mm 3mm 3mm 3mm]{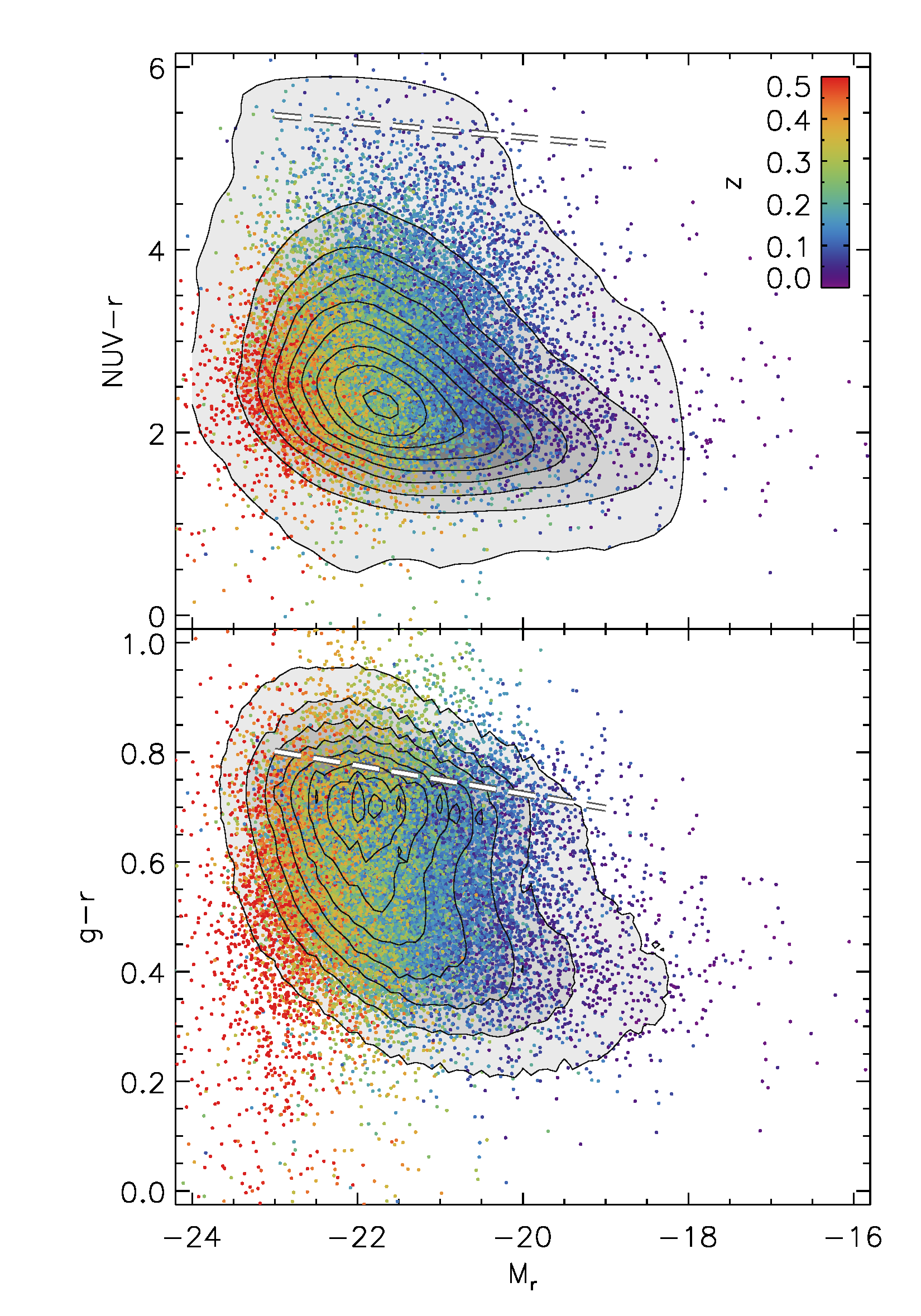}
\caption{Rest-frame colour-magnitude diagrams in $NUV-r$ and $g-r$ for reliable IDs in GAMA, in comparison to the full GAMA sample (greyscale contours). 
The thick dashed line in each panel shows the red sequence derived from the GAMA sample in \citet{Bourne2012}.
Note that \hatlas\ selects galaxies throughout the full locus of GAMA galaxies.
}
\label{fig:cmds}
\end{center}
\end{figure}

\begin{figure}
\begin{center}
\includegraphics[width=0.48\textwidth,clip,trim=9mm 3mm 3mm 3mm]{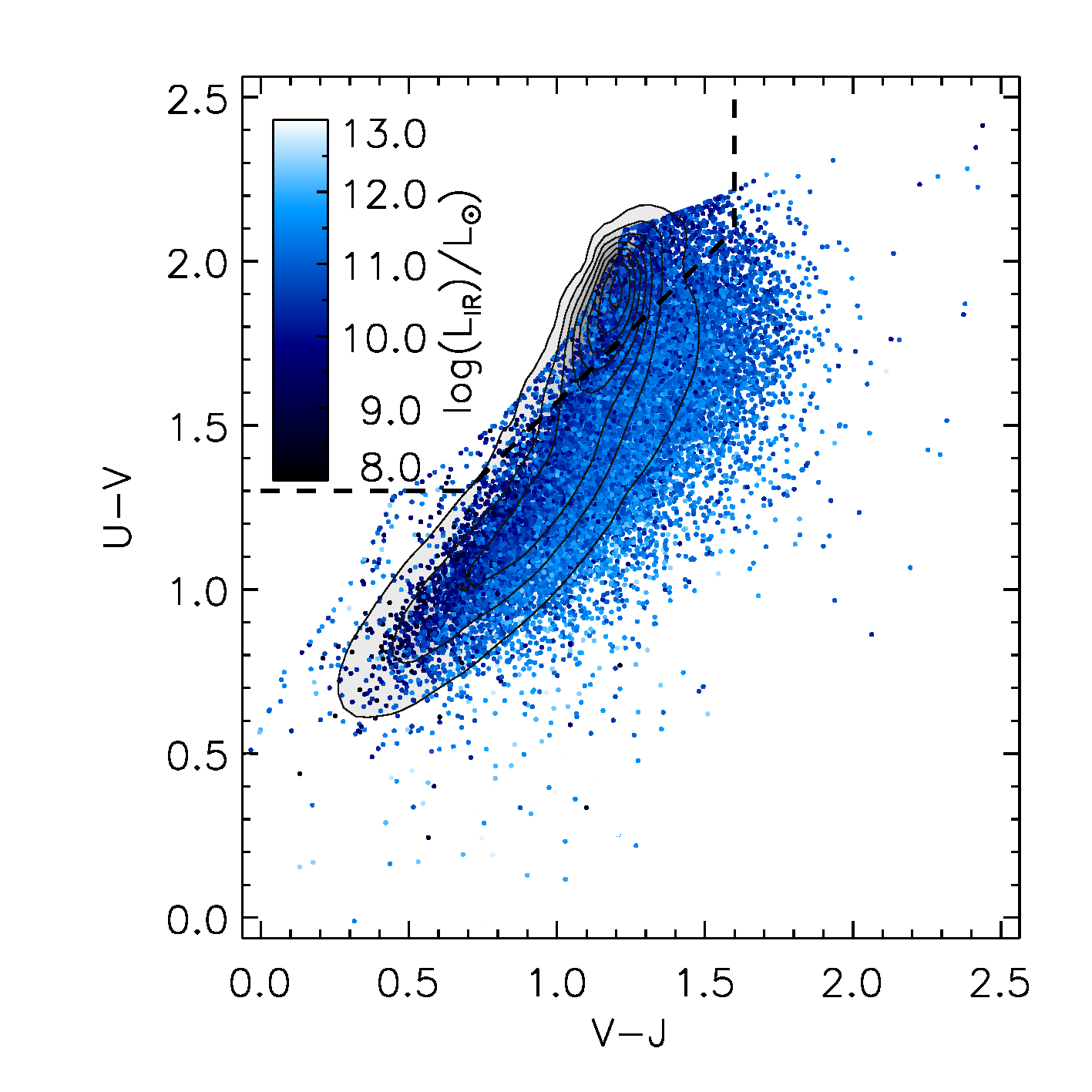}
\caption{Rest-frame $UVJ$ colour-colour diagrams for reliable IDs in GAMA, in comparison to the full GAMA sample (greyscale contours). 
\hatlas\ galaxies are colour-coded by the total IR luminosity estimated from the 100-500\mum\ photometry as described in Section~\ref{sec:lir}.
Note that \hatlas\ selects galaxies throughout the full locus of the GAMA sample, including both passive and star-forming galaxies according to the diagnostic given by the dashed line \citep{Williams2009}.
The sharp edge seen in the top-left locus of the data is artificial and results from the library of model SEDs within \textsc{kcorrect}, which limit the range of colours in these reconstructed photometric bands. 
}
\label{fig:uvj}
\end{center}
\end{figure}

The stellar population age of a galaxy is roughly traced by its optical colour, since quiescent elliptical galaxies are known to reside in the `red sequence' while active star-forming discs are most likely to appear in the `blue cloud'. It has previously been shown \citep[e.g.][]{Dariush2011} that \hatlas\ selects galaxies in both regions of the colour-magnitude diagram, and in Fig.~\ref{fig:cmds} we show that this is indeed the case in the DR1 sample, both in terms of $NUV-r$ and $g-r$ (rest-frame) colours.
The colours have been $k$-corrected to the rest-frame by fitting the $FUV-K$ photometry with {\sc kcorrect} \citep{Blanton2007}.
The redder galaxies in \hatlas\ are likely to be reddened by dust attenuation in the optical \citep{Smith2012c}, although we also know that a small number of truly passive galaxies are detected in the \hatlas\ sample \citep{Rowlands2011,Agius2013}.
We can infer this by predicting the $U-V$ and $V-J$ colours of the reliable IDs, using {\sc kcorrect} to reconstruct the colours in the rest-frame SED. The $UVJ$ diagram in Fig.~\ref{fig:uvj} can then be used to define passive galaxies as those falling within the upper-left region \citep{Williams2009}. Figure~\ref{fig:uvj} shows that a small fraction (8.9 per cent) of sources in the reliable ID sample fall within this region, and these are generally optically luminous. The fraction of the full GAMA sample that falls within the passive region is 35 per cent. 

\begin{figure}
\begin{center}
\includegraphics[width=0.48\textwidth,clip,trim=6mm 3mm 6mm 3mm]{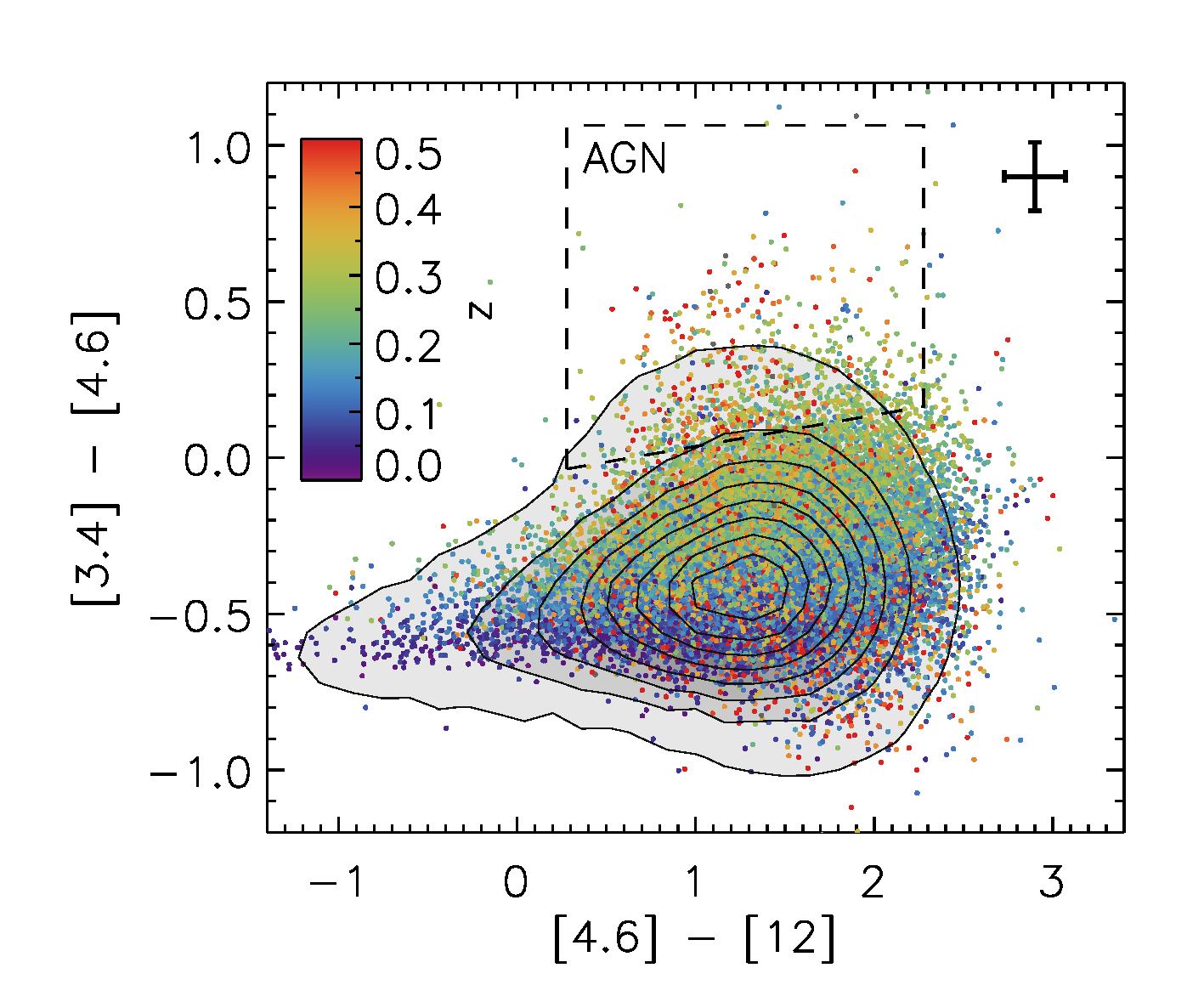}
\caption{A \emph{WISE} colour-colour diagram for reliable \hatlas\ IDs in GAMA, compared with the full GAMA sample (greyscale contours).  Only objects detected in all three bands at SNR~$>3$ are plotted.
The dashed line delineates the locus of AGN following \citet{Jarrett2011}, converted from Vega to AB colours, and the error bars represent the median errors of the \hatlas\ sample.
}
\label{fig:wisecolours}
\end{center}
\end{figure}

Mid-IR colours in the \emph{WISE} bands can be used to classify galaxies according to the emission features resulting from star-formation and dust-obscured active galactic nuclei (AGN). In Fig.~\ref{fig:wisecolours} we show one such classification diagram for galaxies in the reliable ID sample and those in the main GAMA sample that are detected in the $W1$-$W3$ bands.  
The region bounded by the dashed line was defined by \citet{Jarrett2011} to select AGN-dominated mid-IR SEDs, which are generally flat in $\nu L_\nu(\lambda)$ while the SEDs of pure star-forming galaxies are rising in the mid-IR. 
We find that 4.0 per cent of \hatlas\ galaxies are AGN-dominated in the mid-IR, similar to the fraction (6.3 per cent) of GAMA galaxies, although galaxies at higher redshift are clearly more likely to fall within the AGN classification. 
The WISE colours are not $k$-corrected, although \citet{Cluver2014} showed that the $k$-corrected colours in the GAMA sample have a similar distribution to observed colours. Our results are similar to the fraction (5.7 per cent) found in the smaller \hatlas\ SDP sample \citep{Bond2012} using the same diagnostic.

These fractions relate to the subset of galaxies detected in the three \emph{WISE} bands plotted. An alternative classification was proposed by \citet{Messias2012} based on the $K-[4.5]$ and $[4.5]-[8.0]$ colours from \emph{Spitzer}/IRAC photometry, providing a cleaner separation of AGN and star-formation dominated SEDs at a wide range of redshifts.  
For the GAMA/\hatlas\ sample, which is predominantly at redshifts $z<0.5$, a single colour cut of $K-[4.6]>0.2$ should be sufficient to separate SEDs dominated by AGN emission at 4.6\mum\ \citep[see fig. 3 of][]{Messias2012}.
This cut selects 2.7 per cent of \hatlas\ galaxies detected in $K$ and $W2$, compared with 1.9 per cent of GAMA galaxies.  Taking into account upper limits in $W2$, these fractions become $2.7\pm0.1$ per cent for \hatlas\ IDs, and $0.193\pm0.003$ per cent of GAMA galaxies detected in $K$ (errors represent Poisson counting errors only).  

In summary, for the subset of objects detected by $WISE$, the fraction of mid-IR bright AGN in GAMA and \hatlas\ galaxies is similar, but after removing the $WISE$ detection requirement, there is a slightly raised fraction of AGN in \hatlas. This may be interpreted in terms of a link between AGN activity and FIR emission (from star formation or otherwise) but we leave further discussion for dedicated papers.

\subsection{Integrated IR luminosities and obscuration of star formation}
\label{sec:lir}
\begin{figure}
\begin{center}
\includegraphics[width=0.48\textwidth,clip,trim=9mm 3mm 5mm 3mm]{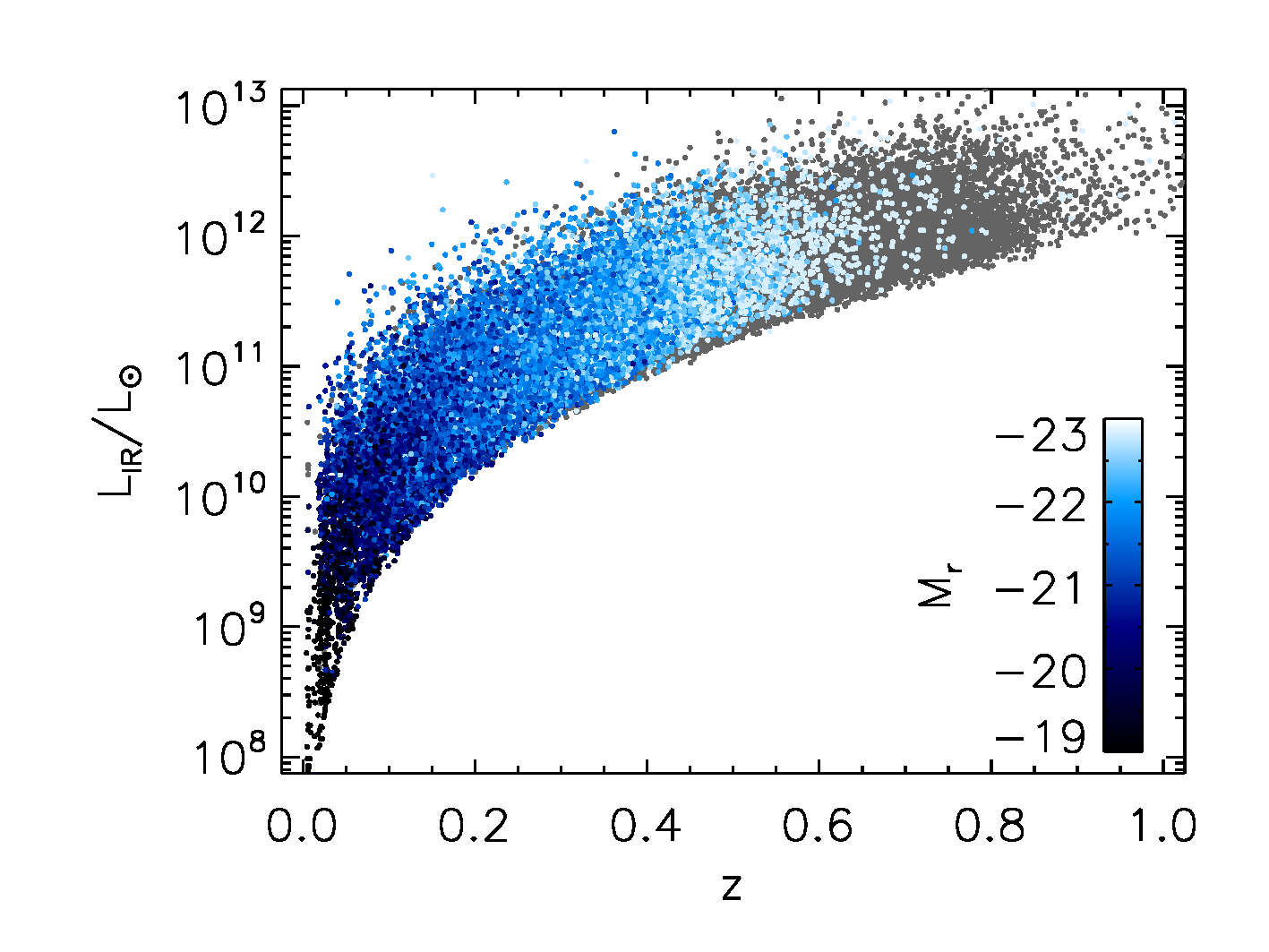}
\caption{Total IR luminosities of all reliable extragalactic \hatlas\ IDs as a function of redshift.
Luminosities were estimated by fitting the 100-500\mum\ fluxes with a modified blackbody SED with fixed $\beta=1.82$ and free temperature and normalization,  
and applying a correction factor of 1.35 as described in the text.
For the subset of IDs in GAMA, we colour the data by the $r$-band absolute magnitude measured from {\sc kcorrect}.}
\label{fig:lir_z}
\end{center}
\end{figure}

We estimated total IR luminosites from the PACS and SPIRE fluxes (100-500\mum) by fitting a single modified blackbody with fixed $\beta=1.82$ \citep{Smith2013}, with temperature and normalization as free parameters, and integrating the model in the range 8--1000\mum.
This model does not account for the warm dust emitting at $\lambda\lesssim100\mum$, and so will under-estimate the total luminosity. A better estimate would be obtained by modelling an additional power-law component in the SED at short wavelengths, which can be constrained by including the WISE 22\mum\ flux \citep{Casey2012}. However, only 14 per cent of our sample are detected at 22\mum, and 9 per cent at 100\mum, so this cannot be constrained for most of the sample. We therefore adopted a single modified blackbody fit to obtain $L_\text{IR}$, but applied a correction factor of 1.35 to all galaxies to account for the average contribution from warm dust \citep[e.g.][]{Casey2012}.
This approximation will under-estimate the total IR luminosity of hotter galaxies, chiefly (ultra)-luminous IR galaxies (ULIRGs, $L_\text{IR}>10^{12}$L$_{\odot}$ and LIRGs, $L_\text{IR}>10^{11}$L$_{\odot}$), but is sufficient to estimate the luminosities of typical star-forming galaxies on average.
Figure~\ref{fig:lir_z} shows the integrated luminosities of the \hatlas\ IDs as a function of redshift. The sample spans a wide range of IR luminosities but becomes dominated by LIRGs
at $z\gtrsim0.25$. 

\begin{figure}
\begin{center}
\includegraphics[width=0.48\textwidth,clip,trim=6mm 3mm 3mm 3mm]{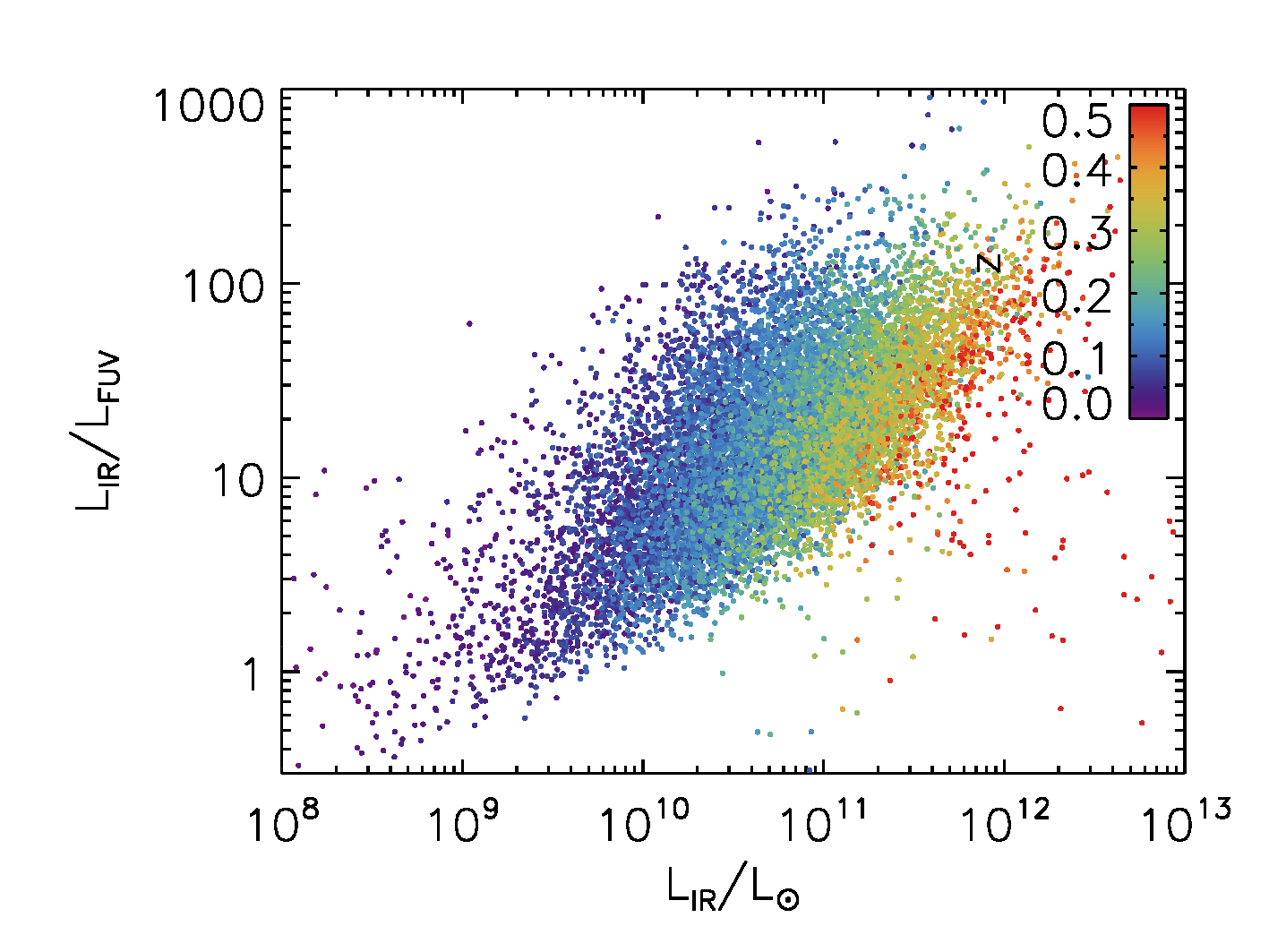}
\includegraphics[width=0.48\textwidth,clip,trim=6mm 3mm 3mm 6mm]{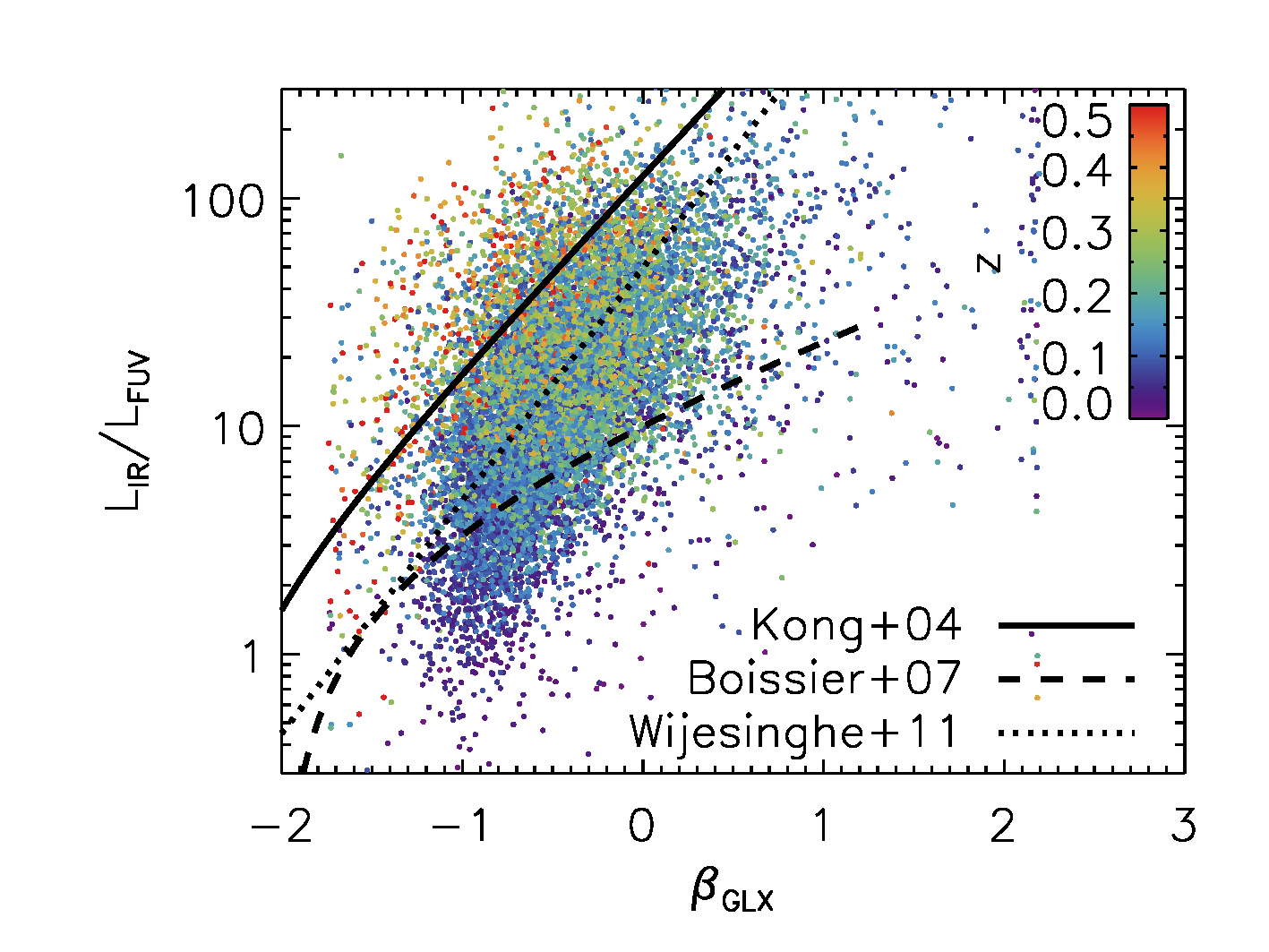}
\caption{Ratio of rest-frame total IR luminosity to rest-frame $FUV$-band luminosity ($\nu L_{\nu}$) for \hatlas\ sources which have reliable IDs in GAMA and $FUV$ detections from \emph{GALEX}, as a function of $L_\text{IR}$ (top) and $\beta_\text{GLX}$ (bottom), where $\beta_\text{GLX}=(\log f_{FUV} - \log f_{NUV})/(\log \lambda_{FUV}-\log \lambda_{NUV})$ as defined by \citet{Kong2004}. Also shown on the lower panel are the best-fit relations for local starbursts \citep{Kong2004}, local late-type galaxies \citep{Boissier2007}, and galaxies selected from the \hatlas\ SDP+GAMA \citep{Wijesinghe2011}. 
}
\label{fig:irx_lir}
\end{center}
\end{figure}

The rest-frame IR-to-UV luminosity ratio was calculated by dividing the integrated IR luminosity (W) by the rest-frame $FUV$ luminosity ($\nu L_{\nu}$/W). 
The range of IR/UV ratios probed by \hatlas\ IDs in GAMA/\emph{GALEX} is similar to that in the \Herschel+\emph{GALEX}-selected sample of \citet{Buat2010}. As expected, it is less biased towards high ratios than \textit{IRAS}-selected LIRGs \citep[e.g.][]{Howell2010}, and less biased towards low ratios than UV-selected samples \citep[e.g.][]{Meurer1999,Kong2004,Overzier2010}.
These other samples have typical log(IR/UV)~$\sim 2$ and $\sim 0$--1 respectively, compared with the median log(IR/UV)~$=0.95$ in the current sample.
The IR/UV ratio traces the relative rates of obscured and unobscured star formation, and in Fig.~\ref{fig:irx_lir} it shows a strong correlation with IR luminosity,
and a weaker correlation with the UV spectral index $\beta$. We use the \emph{GALEX} colour 
$\beta_\text{GLX}=(\log f_{FUV} - \log f_{NUV})/(\log \lambda_{FUV}-\log \lambda_{NUV})$ to estimate $\beta$, following \citet{Kong2004}. The correlation between IR/UV and $\beta$ is well-documented in the literature, and has been shown to depend strongly on the star-formation history, with galaxies dominated by younger stellar populations having higher IR/UV for the same value of $\beta$ \citep{Kong2004,Buat2012}.
The heterogenous \hatlas\ sample therefore shows a large amount of scatter in this relationship. 
At a given value of $\beta_\text{GLX}$, the sample spans a range of IR/UV ratios that encompasses the relationships for starbursts from \citet{Kong2004}, and resolved late-type discs from \citet{Boissier2007}. The best-fit relationship from \citet{Wijesinghe2011}, based on the \hatlas\ SDP+GAMA sample, unsurprisingly provides a good fit to our sample, but clearly a single IR/UV--$\beta$ relationship cannot be assumed without accounting for additional variables within the sample.

\section{Conclusions}
In this paper we have described the process of obtaining reliable optical and multi-wavelength counterparts to submm sources in the \hatlas\ DR1 with UV to near-IR data from SDSS, GAMA and other wide-area surveys. The catalogues described in this paper represent a factor 10 increase in area compared with the previous data release (SDP; \citetalias{Smith2011a}) and an even greater increase in the number of sources due to improvements in the source extraction described in \citetalias{Valiante2014}.

We searched for counterparts within a radius of 10~arcsec around all $\geq4\sigma$ SPIRE sources in the \hatlas\ catalogues, corresponding to an average flux limit of 27.8\,mJy at 250\mum.
For the matching catalogue, we used primary objects from SDSS DR7 and DR9 with $r_\text{model}<22.4$, removing spurious deblends by visual inspection of objects with deblend flags. 
We used likelihood ratios to measure the reliability of all potential matches in this catalogue, and we also inspected by eye a subset of bright sources to validate the automated process and correct missing ID information for bright galaxies and stars.

The intrinsic fraction of \hatlas\ sources which have a counterpart in the SDSS $r<22.4$ catalogue (irrespective of whether the counterpart could be reliably identified) 
is $Q_0=0.539\pm0.001$ overall ($0.519\pm0.001$ for extragalactic objects; $0.020\pm0.002$ for stars).
The extragalactic value is significantly lower than that found by \citetalias{Smith2011a} in the SDP release, but this can be explained by changes in the methodology leading to an improved estimate of $Q_0$.

While $Q_0$ represents the theoretical maximum identification rate for \hatlas\ sources in SDSS, in reality it is not possible to obtain reliable counterparts for this fraction of sources due to the ambiguity of matching submm sources with a PSF of 18~arcsec to optically detected sources, and the poor correlation between the fluxes in these two wavebands.
We found `reliable' counterparts (reliability $R\geq0.8$) for 44,835 submm sources and estimated that this sample has an overall completeness of 73.0 per cent and cleanness of $95.3\pm0.1$ per cent. 
The sample is dominated by galaxies (97 per cent), although 422 stars and 610 spectroscopically identified quasars are also reliably identified, in addition to 351 quasar candidates (unresolved objects without spectroscopy but with non-stellar colours). The stellar IDs are incomplete and uncertain as a result of poor statistical knowledge of the magnitude distribution of submm-emitting stars; further work in this area would improve the efficiency of the LR procedure for stars.
The $R\geq 0.8$ reliability cut excludes some sources with multiple optical counterparts (chance alignments, mergers and pairs/groups with small projected separations can all lead to a completely blended submm source), but we estimate that the incompleteness resulting from the existence of multiple counterparts is small ($\sim 1$ per cent). 

Finally we investigated some of the multi-wavelength properties of \hatlas\ galaxies with counterparts in GAMA. 
We measured the redshift distribution of \hatlas\ reliable IDs, which is broadly distributed, peaking between $0.1<z<0.8$, although there are many \hatlas\ sources that fall below the optical detection limit of SDSS, which are likely to extend to higher redshifts. 
We showed that there is a poor correlation between submm flux and optical magnitude, but that \hatlas\ is sensitive to a wide variety of galaxy types (classified by colour) similar to that probed by GAMA.
For example, \hatlas\ IDs fill the locus occupied by the GAMA colour-magnitude relation in both $g-r$ and $NUV-r$, and about 9 per cent of \hatlas/GAMA matches are classified as passive by their $UVJ$ colours \citep{Williams2009}, compared with 35 per cent of the GAMA sample.  

The \hatlas\ galaxies span five orders of magnitude in integrated IR luminosity at $z<0.5$, and more than two orders of magnitude in IR/UV luminosity ratio. This sample is far less biased than earlier wide-field far-IR-selected samples (e.g. \emph{IRAS}) and represents a valuable census of dust emission throughout the galaxy population at low to intermediate redshifts, as well as many rare objects at high redshifts.

This \hatlas\ data release provides an important legacy of far-IR/submm photometry over a large sky area, and the multi-wavelength catalogue described in this paper has unrivalled potential for studying the interstellar medium and obscured star formation in a large sample of galaxies up to redshift $z\lesssim1$.
All of the data described in this paper and in \citetalias{Valiante2014} are available from \url{www.h-atlas.org}.

\section*{Acknowledgements}
We would like to thank Micha{\l} Micha{\l}owski, Hugo Messias and Myrto Symeonidis for advice and enlightening discussions while preparing this paper.
NB has received funding from the European Union Seventh Framework Programme (FP7/2007-2013) under grant agreement no. 312725;
SD is supported by STFC with an Ernest Rutherford fellowship; and
CF acknowledges funding from CAPES (proc. 12203-1).
LD, SJM and RJI acknowledge support from European Research Council Advanced Investigator Grant COSMICISM; LD and SJM are also supported by consolidated grant COSMICDUST.
SE and EV thank the UK Science and Technology Facilities Council for financial support.
MS and SE have received funding from the European Union Seventh Framework Programme ([FP7/2007-2013] [FP7/2007-2011]) under grant agreement no. 607254.

For the least-squares fitting in this work we have made use of the {\sc mpfit} package, available from \url{http://purl.com/net/mpfit} \citep{Markwardt2009}.
The \Herschel-ATLAS is a project with \Herschel, which is an ESA space observatory with science instruments provided by European-led Principal Investigator consortia and with important participation from NASA. The \hatlas\ website is \url{http://www.h-atlas.org/}.
GAMA is a joint European-Australasian project based around a spectroscopic campaign using the Anglo-Australian Telescope. The GAMA input catalogue is based on data taken from the Sloan Digital Sky Survey and the UKIRT Infrared Deep Sky Survey. Complementary imaging of the GAMA regions is being obtained by a number of independent survey programs including \emph{GALEX} MIS, VST KiDS, VISTA VIKING, \emph{WISE}, \hatlas, GMRT and ASKAP providing UV to radio coverage. GAMA is funded by the STFC (UK), the ARC (Australia), the AAO, and the participating institutions. The GAMA website is \url{http://www.gama-survey.org/}.

Funding for the SDSS, SDSS-II and SDSS-III has been provided by the Alfred P. Sloan Foundation, the Participating Institutions, the National Science Foundation, the U.S. Department of Energy, the National Aeronautics and Space Administration, the Japanese Monbukagakusho, the Max Planck Society, and the Higher Education Funding Council for England. The SDSS Web Site is \url{http://www.sdss.org/}.
The SDSS is managed by the Astrophysical Research Consortium for the Participating Institutions. The Participating Institutions are the American Museum of Natural History, Astrophysical Institute Potsdam, University of Basel, University of Cambridge, Case Western Reserve University, University of Chicago, Drexel University, Fermilab, the Institute for Advanced Study, the Japan Participation Group, Johns Hopkins University, the Joint Institute for Nuclear Astrophysics, the Kavli Institute for Particle Astrophysics and Cosmology, the Korean Scientist Group, the Chinese Academy of Sciences (LAMOST), Los Alamos National Laboratory, the Max-Planck-Institute for Astronomy (MPIA), the Max-Planck-Institute for Astrophysics (MPA), New Mexico State University, Ohio State University, University of Pittsburgh, University of Portsmouth, Princeton University, the United States Naval Observatory, and the University of Washington.

\bibliographystyle{mn2e}
\bibliography{Bib_p1ids}
\bsp 
\label{lastpage}
\end{document}